\documentclass[twocolumn]{aastex62}
\usepackage{ulem,xcolor}

\newcommand{\Msun}{\mbox{\,$M_{\odot}$}}

\usepackage{epsf}
\usepackage{graphicx}
\usepackage{natbib}
\usepackage{url}
\usepackage{mathrsfs}
\usepackage{amsmath}
\font\smcap=cmcsc10

\newcommand{\kms}{\,km~s$^{-1}$}

\newcommand{\nai}{Na\,{\smcap i}}
\newcommand{\caii}{Ca\,{\smcap ii}}
\newcommand{\feh}{$\rm[Fe/H]$}
\newcommand{\afe}{$\rm[\alpha/Fe]$}
\newcommand{\teff}{$T_{\rm eff}$}
\newcommand{\fehp}{$\rm[Fe/H]_{phot}$}
\newcommand{\fehsynth}{$\rm[Fe/H]_{synth}$}
\newcommand{\meanfehsynth}{$\langle$[Fe/H]$\rangle_{\rm synth}$}
\newcommand{\meanafe}{$\langle$[$\alpha$/Fe]$\rangle$}
\newcommand{\fehcat}{$\rm[Fe/H]_{CaT}$}

\newcommand{\rproj}{$R_{\rm proj}$}
\newcommand{\vio}{$(V-I)_0$}
\newcommand{\ivi}{($I,\,V-I$)}

\shorttitle{Alpha and Iron Abundances in M31's Outer Halo}
\shortauthors{Gilbert et al.}

\newcommand{\nouterhalofeh}{23}
\newcommand{\nouterhalofehours}{21}
\newcommand{\nAndIhalofeh}{7}

\newcommand{\nouterhaloafe}{9}
\newcommand{\nouterhaloafeours}{7}
\newcommand{\nbootstrapdraws}{$10^4$}

\newcommand{\minrprojinnerhalo}{12}
\newcommand{\maxrprojinnerhalo}{26}
\newcommand{\minrprojafeouterhalo}{45}
\newcommand{\maxrprojafeouterhalo}{150} % including Luis' two stars
\newcommand{\minrprojfehouterhalo}{43}
\newcommand{\maxrprojfehouterhalo}{165}

\newcommand{\rprojdisk}{$26$} 
\newcommand{\rprojkmggss}{$17$}  %f207
\newcommand{\rprojH}{$12$} 
\newcommand{\rprojS}{$22$} 
\newcommand{\rprojiehalo}{$23$}  %f130
\newcommand{\rprojandone}{$45$} % mean for the deep AndI masks

\newcommand{\meanafepastseventy}{$0.33\pm0.17$}

\newcommand{\meanafeouter}{$0.30\pm0.16$} % all
\newcommand{\meanafeinner}{$0.45\pm0.09$} % halo component only
\newcommand{\weightedmeanafeinner}{$0.44$} % probabilistic halo component ; weighted by P_halo and measurement uncertainty
\newcommand{\meanafeinnerfehrestricted}{$0.47\pm0.13$} % halo component only, restricted to feh < -1.1
\newcommand{\meanfehuncertainties}{0.2}

\newcommand{\meanfehandIall}{$-1.27\pm0.19$}

\newcommand{\meanfehandIexcltwosigma}{$-1.60\pm0.11$}

\newcommand{\andIphotbias}{$0.55$}
\newcommand{\andIfehphotselection}{$-0.95$}

\begin{document}

\title{Elemental Abundances in M31: Iron and Alpha Element Abundances in M31's Outer Halo\footnote{The data presented herein were obtained at the W.M. Keck Observatory,
which is operated as a scientific partnership among the California
 Institute of Technology, the University of California, and the National
Aeronautics and Space Administration. The Observatory was made
possible by the generous financial support of the W.M. Keck
Foundation.}}

\correspondingauthor{Karoline M. Gilbert}
\email{kgilbert@stsci.edu}

\author[0000-0003-0394-8377]{Karoline M. Gilbert}
\affiliation{Space Telescope Science Institute, 3700 San Martin Dr., Baltimore, MD 21218, USA}
\affiliation{Department of Physics \& Astronomy, Bloomberg Center for Physics and Astronomy, Johns Hopkins University, 3400 N. Charles Street, Baltimore, MD 21218, USA}

\author[0000-0002-3233-3032]{Jennifer Wojno}
\affiliation{Department of Physics \& Astronomy, Bloomberg Center for Physics and Astronomy, Johns Hopkins University, 3400 N. Charles Street, Baltimore, MD 21218, USA}

\author[0000-0001-6196-5162]{Evan N. Kirby}
\affiliation{California Institute of Technology, 1200 E. California Boulevard, MC 249-17, Pasadena, CA 91125, USA}

\author[0000-0002-9933-9551]{Ivanna Escala}
\affiliation{California Institute of Technology, 1200 E. California Boulevard, MC 249-17, Pasadena, CA 91125, USA}
\affiliation{Department of Astrophysical Sciences, Princeton University, 4 Ivy Lane, Princeton, NJ, 08544, USA}

\author[0000-0002-1691-8217]{Rachael L. Beaton}
\altaffiliation{Hubble Fellow.}
\altaffiliation{Carnegie-Princeton Fellow.}
\affiliation{Department of Astrophysical Sciences, Princeton University, 4 Ivy Lane, Princeton, NJ, 08544, USA}
\affiliation{The Observatories of the Carnegie Institution for Science, 813 Santa Barbara St., Pasadena, CA, 91101, USA}

\author[0000-0001-8867-4234]{Puragra Guhathakurta}
\affiliation{UCO/Lick Observatory, Department of Astronomy \& Astrophysics, University of California Santa Cruz, 
 1156 High Street, 
 Santa Cruz, CA, 95064, USA}

\author[0000-0003-2025-3147]{Steven R. Majewski}
\affiliation{Department of Astronomy, University of Virginia, Charlottesville, VA 22904-4325, USA}

\begin{abstract}

We present \feh\ and \afe\ abundances, derived using spectral synthesis techniques, for stars in M31's outer stellar halo.  The \nouterhalofehours\ \feh\ measurements and \nouterhaloafeours\ \afe\ measurements are drawn from fields ranging from \minrprojfehouterhalo\ to \maxrprojfehouterhalo~kpc in projected distance from M31.  We combine our measurements with existing literature measurements, and compare the resulting sample of \nouterhalofeh\ stars with \feh\ and \nouterhaloafe\ stars with \afe\ measurements in M31's outer halo with \afe\ and \feh\ measurements, also derived from spectral synthesis, in M31's inner stellar halo ($r < $\,\maxrprojinnerhalo~kpc) and dSph galaxies.  The stars in M31's outer halo have \afe\ patterns that are consistent with the largest of M31's dSph satellites (And~I and And~VII).  These abundances provide tentative evidence that the \afe\ abundances of stars in M31's outer halo are more similar to the abundances of Milky Way halo stars than to the abundances of stars in M31's inner halo. We also compare the spectral synthesis-based \feh\ measurements of stars in M31's halo with previous photometric \feh\ estimates, as a function of projected distance from M31.  The spectral synthesis-based \feh\ measurements are consistent with a large-scale metallicity gradient previously observed in M31's stellar halo to projected distances as large as 100~kpc.      

\end{abstract}


\section{Introduction}\label{sec:intro}

The merger history of a galaxy is encoded in its stellar halo population, which is thought to be built primarily from the accretion of smaller systems \citep[e.g.,][]{bullock2005}.  Recent work suggests that the mean metallicity of a stellar halo is primarily determined by the mass of the dominant halo progenitor \citep{dsouza2018MNRAS,dsouza2018} or progenitors \citep{monachesi2019}, with the implication that a relatively simple measurement can provide meaningful observational constraints on the most significant merger event in a galaxy's history.  Gradients in metallicity over the extent of a stellar halo can also provide observational constraints on the mass of the dominant progenitor as well as the relative masses of the progenitors of the inner and outer regions of a stellar halo \citep[e.g.,][]{cooper2010,font2011,tissera2013,tissera2014,monachesi2019}, yielding insight into the mass profile of accretion events. 

However, additional information is needed to further constrain the nature of the progenitors of a galaxy's stellar halo, as well as the timescales over which these accretion events occurred.  The abundances of $\alpha$-elements (O, Ne, Mg, Si, S, Ar, Ca, and Ti) as a function of \feh\ provide observational constraints on the timescale and efficiency of star formation, and thus insights into the nature of the progenitor as well as the time of its accretion into the stellar halo \citep[e.g.,][]{shetrone2001,venn2004, robertson2005,font2006,johnston2008,hayes2018,nidever2020}. 

Measurements of the metallicity of the stellar halo of the large spiral galaxy Andromeda (M31) have largely been derived using resolved stellar photometry via a comparison of the colors and magnitudes of stars with theoretical isochrones (\fehp), which generally requires assuming a fixed age, distance, and \afe\ abundance ratio for the population.  This is the case both for photometric surveys of M31's halo \citep[e.g.,][]{ibata2014} as well as most spectroscopic studies, wherein the spectra were used to identify individual stars as M31 halo members rather than foreground Milky Way (MW) stars or M31 dSph members \citep[e.g.,][]{gilbert2014}.  On average, M31's halo is significantly more metal-rich than that of the MW, and is also metal-rich compared to other nearby galaxies (or simulated galaxies) of roughly similar mass and morphology \citep{monachesi2016,harmsen2017,dsouza2018MNRAS}. This indicates M31's stellar halo likely had a dominant progenitor (or progenitors) more massive than the typical dominant progenitors of MW- and M31-mass galaxies.  

M31's halo exhibits a significant gradient in metallicity as a function of radius, with the mean metallicity of the halo decreasing by $\sim 1$~dex over $\sim 100$~kpc, as shown in global analyses of M31's halo using spectroscopic \citep{gilbert2014} and photometric \citep{ibata2014} datasets (see \citet{gilbert2014} for a review of earlier work).  However, each study assumed both a single age and \afe\ for the stellar population.  This assumption is known to be overly simplistic for M31's inner halo, which exhibits a range of stellar ages in fields extending to projected distances of at least 35~kpc \citep[e.g.,][]{brown2006,brown2008}. Thus, photometrically derived \feh\ distributions in M31's halo are likely to differ from the true \feh\ distribution, with different age or \afe\ assumptions affecting the \feh\ estimates by $\lesssim \pm 0.2$~dex \citep{gilbert2014}.  If a gradient in mean age or \afe\ as a function of radius exists in M31's halo, M31's metallicity gradient may be steeper (or shallower) than measured from the photometric \feh\ estimates \citep{gilbert2014}.    

The equivalent width of the \caii\ triplet absorption feature provides an independent estimate of \feh\ (\fehcat) for comparison to \fehp\ \citep[e.g.,][]{armandroff1991,battaglia2008,starkenburg2010,dacosta2016}.  However, estimates of \feh\ based on \caii\ equivalent width measurements in spectroscopic studies of M31 halo stars typically have large uncertainties \citep[e.g.,][]{gilbert2014}, due to a combination of relatively low spectral signal-to-noise ratio (S/N) and limited precision resulting from measuring the equivalent width of only two or three absorption lines that are frequently affected by night sky lines at the velocities typical of M31 halo stars. This has driven the general reliance on photometric-based \feh\ estimates in M31's halo,  
although \fehcat\ was measured by \citet{kalirai2006halo}, \citet{koch2008}, and \citet{gilbert2014}.  Nevertheless, good agreement was found between the strong trends of mean \feh\ as function of projected distance from M31's center by \citet{gilbert2014}, regardless of whether \feh\ was derived from photometric estimates or the strength of the \caii\ triplet absorption feature.  Still, the \caii\ triplet is able to provide only an estimate of metallicity, not an abundance ratio like \afe.

Spectral synthesis techniques compare the observed spectra to a synthetic spectral grid covering a range of effective temperatures, surface gravities, and \feh\ and \afe\ abundances.  Unlike the photometric and \caii\ triplet-based techniques, the spectral synthesis technique enables measurements of both \feh\ and \afe, using regions of the spectrum sensitive to the presence of \feh\ and $\alpha$-elements.  These measurements provide insight into the timescales and efficiency of star formation in the progenitors of M31's outer halo.  Moreover, robustly determining both the mean \feh\ and \afe\ abundance of stars in M31's halo will yield important empirical constraints that can be leveraged to improve the accuracy of photometric \feh\ estimates over large, contiguous areas of M31's halo.  

Initial abundance work in M31 applied this technique to fields targeting M31's dSph satellites \citep{vargas2014}.  Recently, we have applied this technique to deep observations in fields targeting M31's inner halo, as well as deep observations of dSph satellites. 
We have extended the spectral synthesis technique to lower-resolution spectra, making the first measurements of \afe\ for stars in M31's inner halo \citep{escala2019}.  We have measured \afe\ as a function of \feh\ for inner halo fields targeting tidal debris features as well as relatively smooth halo fields without identified substructure. The \afe\ and \feh\ abundances provide strong evidence that the progenitors of both the kinematically hot inner halo and M31's giant stellar stream had significantly different timescales and efficiencies of star formation than those of M31's surviving dSph satellites \citep{gilbert2019,escala2020}, 
indicating M31's inner halo was built primarily from galaxies more massive than M31's surviving dSph satellites. We have also significantly increased the sample of \afe\ and \feh\ measurements in M31's dSph satellites, finding that the trends of [$\alpha$/Fe] with [Fe/H] in M31's satellites depend on the galaxies' stellar masses in the same way as MW satellites \citep{kirby2020}.  We also reported the first \afe\ measurements for several of M31's lower-mass satellites \citep{wojno2020}.

The above studies present significant progress in our understanding of the chemical evolution of the progenitors of M31's stellar halo and dSph satellites.  However, the low density of M31's outer halo makes compiling a sample of M31 outer halo stars with \feh\ and \afe\ abundances derived from spectral synthesis techniques challenging.  \citet{vargas2014ApJL} presented the first---and to date, only---measurements of \afe\ as well as \feh\ for stars in M31's outer halo.  
The four outer halo stars in the \citet{vargas2014ApJL} sample  were observed in spectroscopic fields designed to target the dSph galaxies And~II (144~kpc in projection from M31's center), And~III (68~kpc; two stars), and AndXIII (117~kpc).  All four stars are relatively metal-poor, with $-2.2< {\rm [Fe/H]} <-1.4$.  \citeauthor{vargas2014ApJL}\ found that these outer halo stars have mild $\alpha$-enhancement, with a mean of $\sim+0.2\pm0.2$~dex

In this contribution, we increase the sample of stars in M31's outer halo with \afe\ and \feh\ measurements derived using spectral synthesis techniques. We present \afe\ measurements for seven stars, and \feh\ measurements for 21 stars, in M31's outer halo.  
Two of the stars, both with \afe\ measurements, were also measured in the \citet{vargas2014ApJL} sample.  Section~\ref{sec:data} describes the spectroscopic datasets from which the M31 outer halo sample is drawn, the method for measuring \afe\ and \feh, the quality criteria applied to determine whether the abundance measurements are robust, and the criteria used to isolate M31 halo stars from dSph members and foreground MW stars.  Section~\ref{sec:halo_abundances} presents the distribution of \afe\ with \feh\ of stars in M31's outer halo, and compares it to that of stars in M31's inner halo, dSphs, and a sample of MW halo stars. Section~\ref{sec:cmdfeh} presents the \feh\ distribution of M31 outer halo stars, and compares the \feh\ abundance distributions derived from spectral synthesis with those derived from photometry.  
Section~\ref{sec:conclusion} summarizes our conclusions.

All quoted distances in kpc from M31's center refer to the projected distance in the sky tangent plane (\rproj), assuming a distance modulus of $24^{\rm m}.47$ to M31 \citep{stanek1998, mcconnachie2005}. 
This allows direct comparisons with earlier literature results.  However, we note that this distance is slightly greater than the more recent distance modulus estimate of $24^{\rm m}.38$ based on measurements of Cepheid variables \citep{riess2012}.  Use of the M31 distance based on Cepheid measurements would result in a $\sim4$\% reduction in the estimated \rproj.

\section{M31 Outer Halo Sample Selection and Abundance Measurements }\label{sec:data}

\subsection{Spectroscopic Data Sets}

\begin{figure}[tb]
\includegraphics[width=0.48\textwidth]{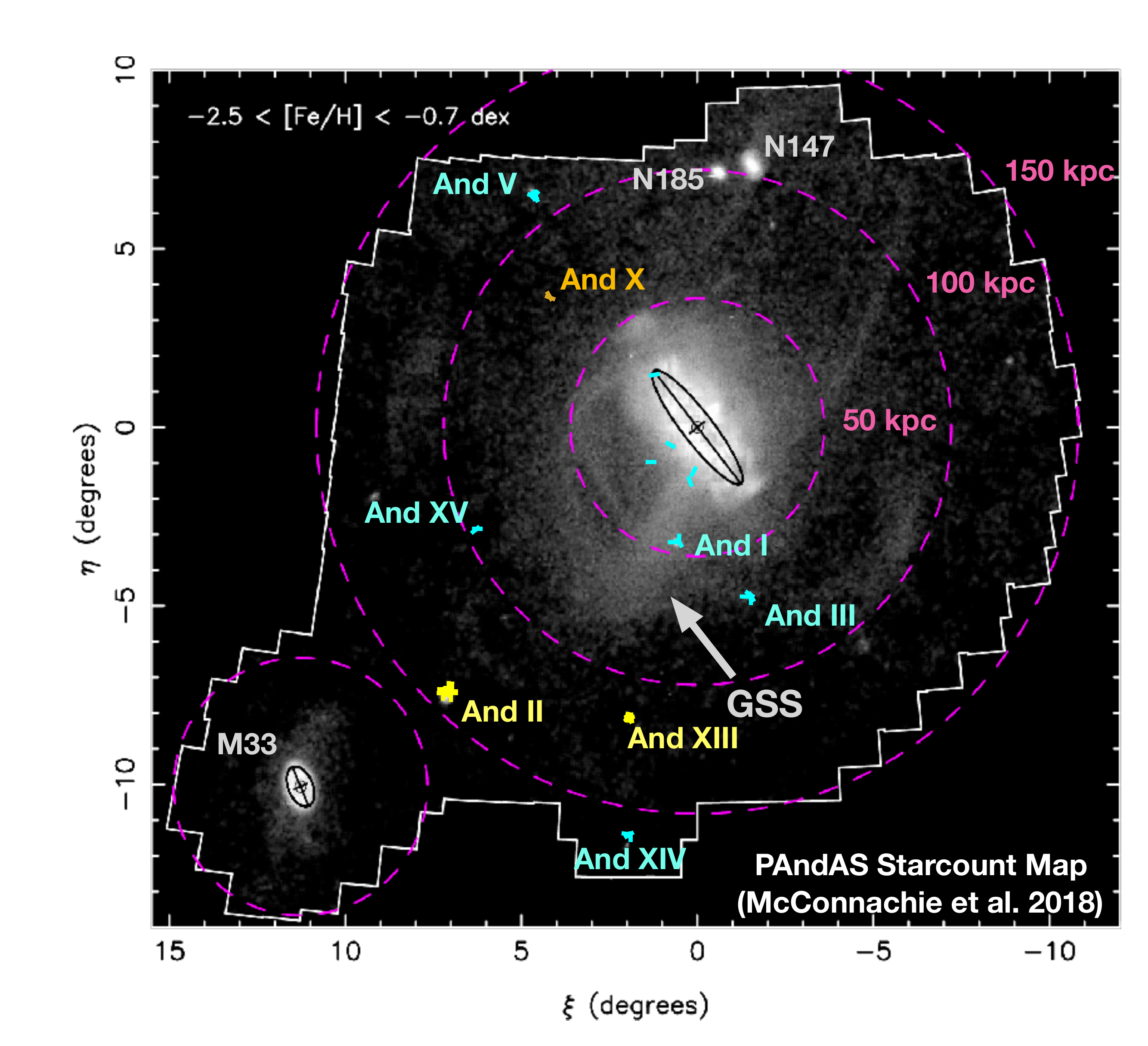}
\caption{Locations of spectroscopic fields with spectral synthesis--based abundance measurements that are discussed in this work.  Fields containing M31 halo stars with \fehsynth\ or \afe\ measurements made in the course of our M31 abundance survey are labeled in cyan.  Fields with M31 halo stars with \fehsynth\ or \afe\ measurements drawn from the literature \citep{vargas2014ApJL} are labeled in yellow.  All spectroscopic masks that were included in the analysis are shown, regardless of whether they yielded M31 halo star abundance measurements passing all quality criteria (Sections~\ref{sec:halo_selection} and ~\ref{sec:abund_measurements}).  While the spectroscopic masks targeting the dSphs And~VII and And~X did not yield any M31 halo star candidates with abundance measurements, we compare the M31 halo star abundances to the abundances of members of And~VII (located off the figure) and And~X (shown in orange). The underlying figure shows the surface density of M31 RGB stars (with $-2.5 <$\fehp\,$<-0.7$, estimated from comparison of the stars' colors and magnitudes with theoretical isochrones) from the PAndAS survey  \citep[Figure~11 of][]{mcconnachie2018}.
}
\label{fig:roadmap}
\end{figure}

The abundance measurements are derived from spectra obtained with the DEIMOS spectrograph \citep{faber2003} on the Keck~II 10~m telescope (Figure~\ref{fig:roadmap}).  The M31 outer halo sample  
presented here is drawn from spectroscopic masks designed to target M31 dSphs, and includes  
both deep and shallow observations obtained with the 1200 line mm$^{-1}$ (1200G) grating ($R\sim 6000$, $\lambda\lambda\sim 6300-9100$\,\AA).  The spectra of the targeted dSph members were analyzed by \citet{kirby2020} and \citet{wojno2020}. 
The deep M31 observations, with total exposure times of approximately 6 hours per mask (slitmasks ``and1a'', ``and3a'', ``and5b''), are described by 
\citet{kirby2020}.
The shallow observations, with typical total exposure times of approximately 1 hour per mask (slitmasks ``d1\_1'', ``d1\_2'', ``d3\_2'', ``d3\_3'', ``d5\_1'', ``d14\_3'', ``d15\_1'', ``d15\_2''), were obtained as part of the SPLASH survey \citep{gilbert2012,tollerud2012} and are   
described by \citet{wojno2020}, who give details about the data reduction that are relevant to the abundance measurements.  All spectra were reduced using the {\tt spec2d} and {\tt spec1d} software \citep{cooper2012,newman2013}, with modifications tailored to stellar spectra as described by \citet{kirby2015a} (for the deep observations) and \citet{gilbert2012} (for the shallow observations).   

\subsection{Selection of M31 Halo Stars}\label{sec:halo_selection}

\begin{figure*}[tbh]
\includegraphics[width=\textwidth]{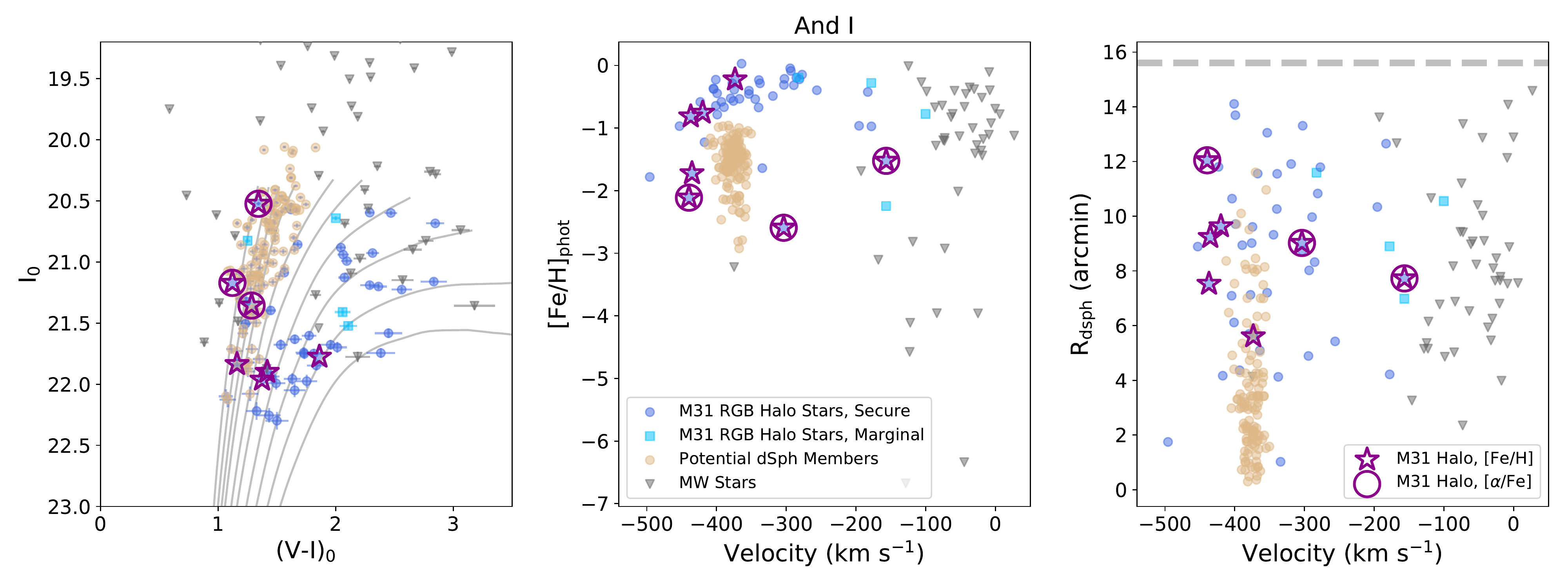}
\includegraphics[width=\textwidth]{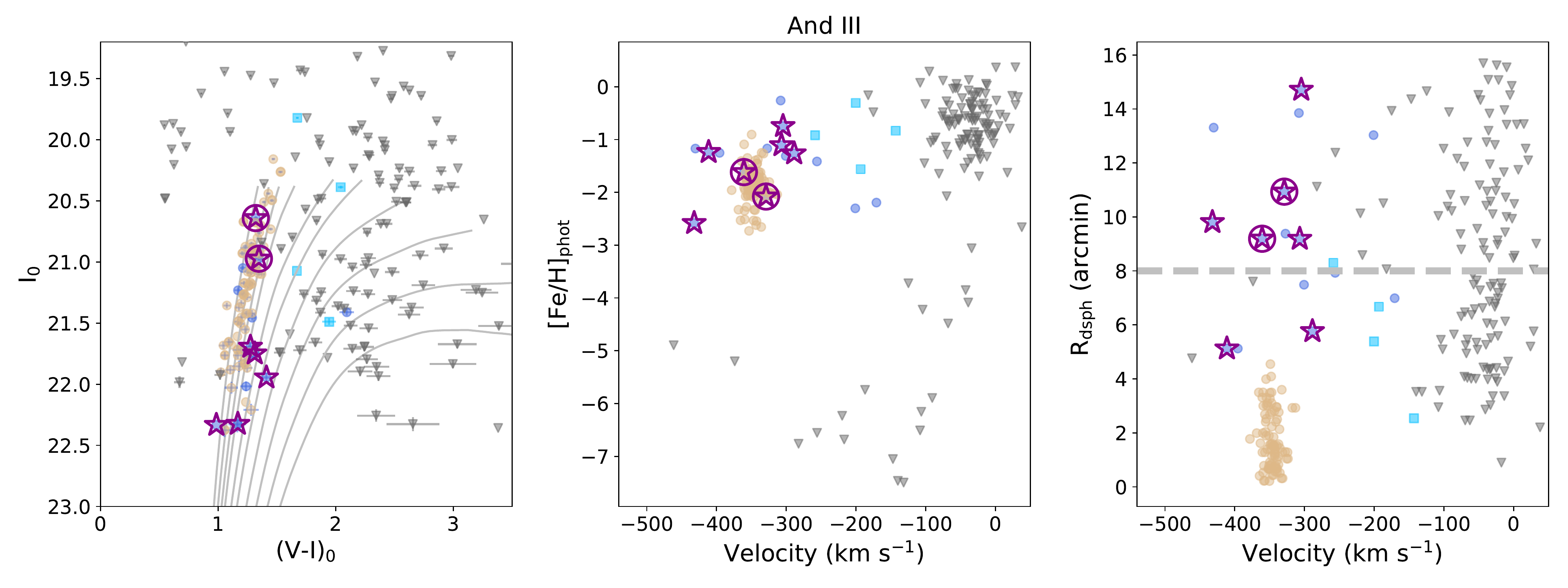}
\caption{Selection of the M31 halo star abundance samples (Section~\ref{sec:data}) in the And~I and And~III fields.   
To be considered in the M31 halo sample, stars were required to be securely classified as RGB stars at the distance of M31 \citep{gilbert2006} and to 
pass cuts in either velocity, projected distance from the dSph, or (for the And~I field only) \fehp\ (Section~\ref{sec:halo_selection}). Stars that pass the M31 halo star selection criteria are denoted with blue points (circles denote stars that are $\geq 3$ times more likely to be RGB stars at the distance of M31 than MW dwarf stars, while squares are $< 3$ times as likely to be M31 stars).  Stars that are classified as red giants but do not pass the halo star selection criteria are considered potential dSph members and are denoted with tan points.
M31 halo stars with \fehsynth\ measurements that pass all quality criteria are shown by the large purple stars, while stars with \afe\ measurements that pass all quality criteria are additionally denoted with large purple circles (Section~\ref{sec:abund_measurements}).  The RGB classification shown here assumes M31's distance modulus and a velocity distribution appropriate for M31 halo stars, and thus is not optimized for classifying RGB stars in the dSphs.  
Isochrones shown in the leftmost panels assume an age of 12 Gyr, \afe$=0$, a distance modulus of 24.$^m$47, and range from \feh\ $=-2.31$ to 0~dex \citep{vandenberg2006}.  Dashed lines in the rightmost panels show four times the half-light radius from the center of the dSph \citep{martin2016}.
}
\label{fig:sampleselection1}
\end{figure*}

\begin{figure*}[tbh]
\includegraphics[width=\textwidth]{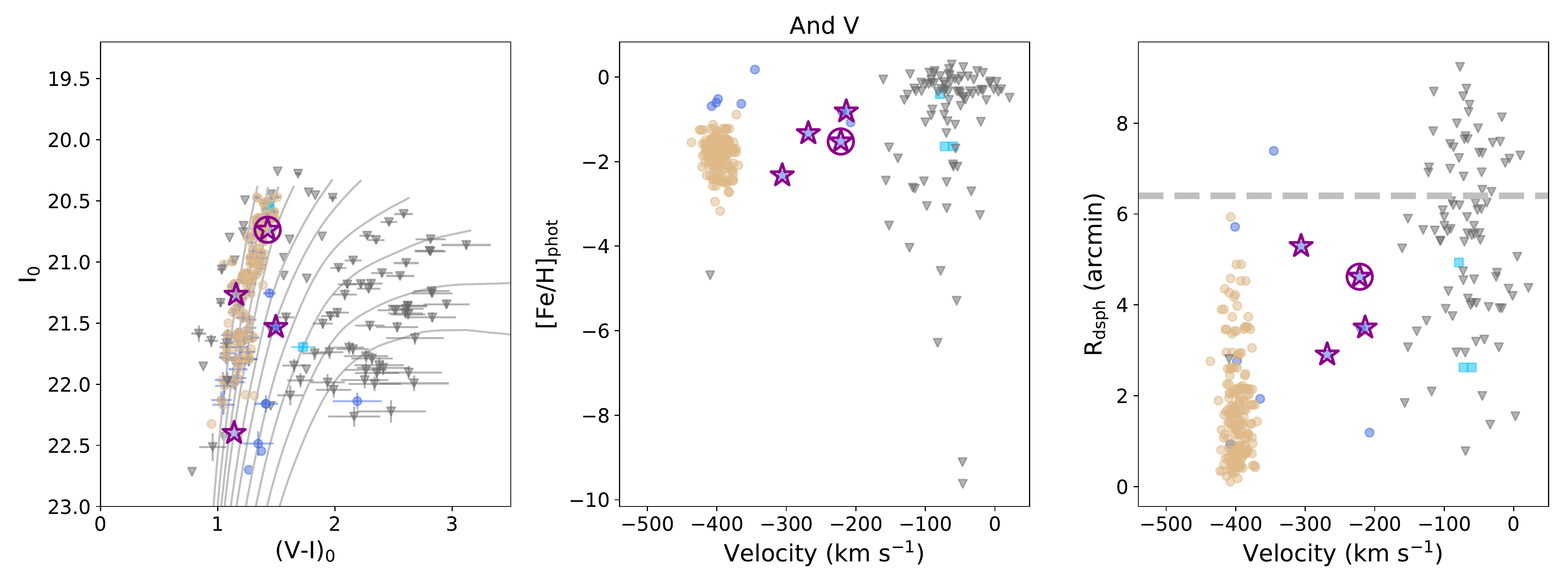}
\includegraphics[width=\textwidth]{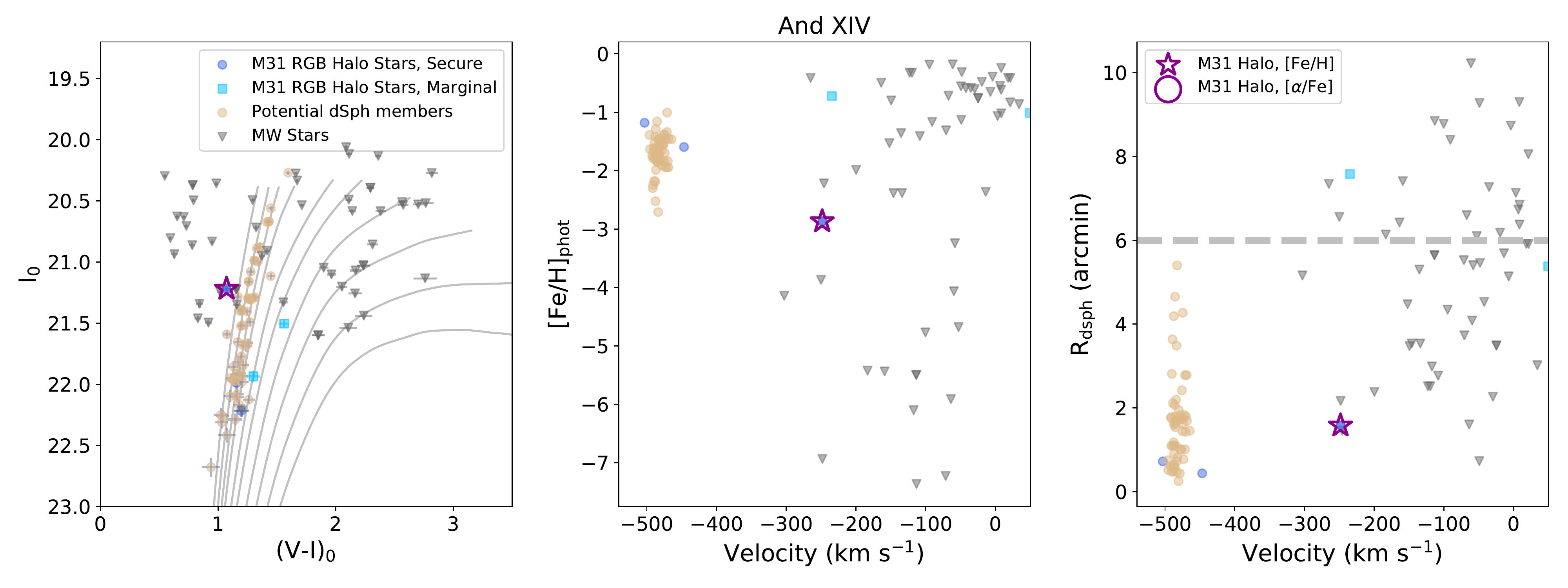}
\includegraphics[width=\textwidth]{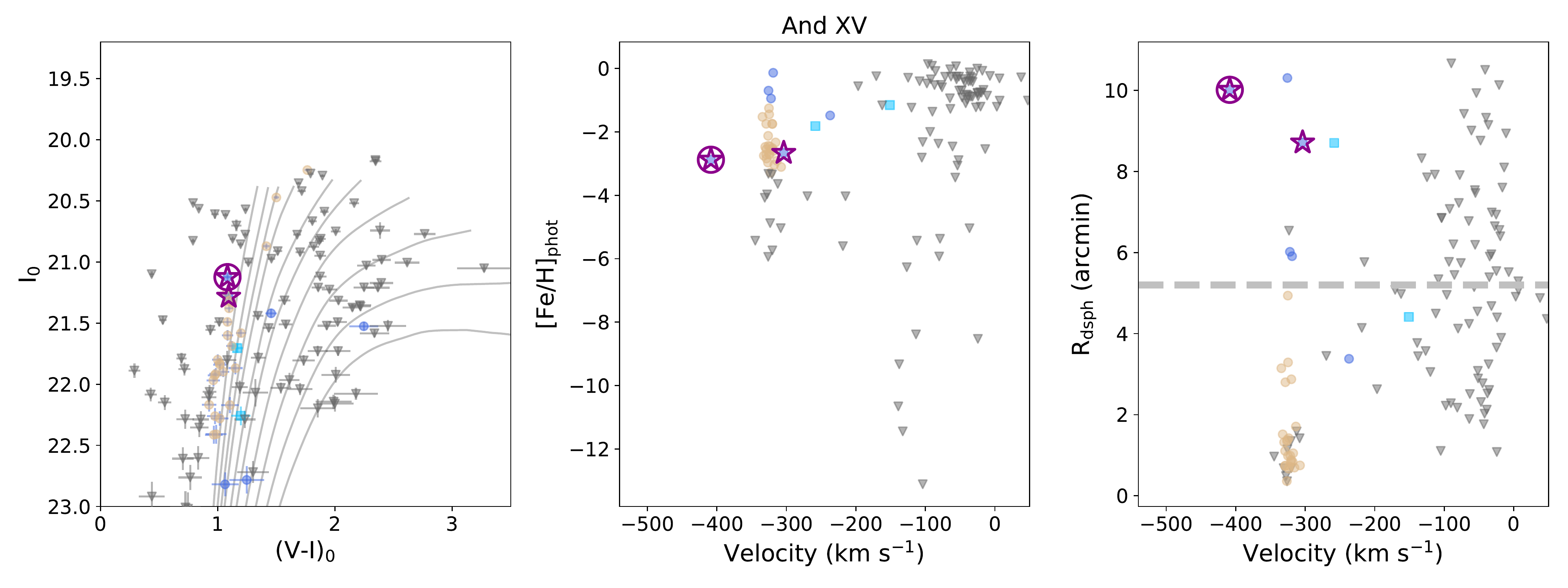}
\caption{Same as Figure~\ref{fig:sampleselection1} for fields And~V, And~XIV and And~XV. 
}
\label{fig:sampleselection2}
\end{figure*}

Stars are identified as probable red giant branch stars at the distance of M31, rather than MW stars along the line of sight, using the empirical diagnostics of \citet{gilbert2006}. These diagnostics include line of sight velocity, surface gravity-sensitive narrowband imaging, position in the color-magnitude diagram, measurement of a surface gravity-sensitive absorption feature (\nai), and a comparison of photometric (\fehp) and spectroscopic (\fehcat) metallicity estimates.  For the sample presented here, we require that stars be at least three times more likely to be RGB stars at the distance of M31 than MW stars. This method is summarized, and the resulting stellar classification is investigated for signs of contamination in the outer halo, by \citet{gilbert2012}.  \citeauthor{gilbert2012}\ conclude that the sample of securely identified M31 outer halo stars are not significantly contaminated by misidentified MW stars. 

All of the outer halo fields considered here targeted dSphs in M31's outer halo. Thus, M31 halo stars must also be distinguished from probable dSph members (Figures~\ref{fig:sampleselection1} and \ref{fig:sampleselection2}).  In previous work \citep[e.g.,][]{gilbert2012,gilbert2014,gilbert2018}, we classified M31 halo stars in dSph fields using a combination of the velocity of the star compared to the mean velocity of the dSph, the projected distance of the star from the center of the dSph compared to estimates of the dSph's tidal radius, and the position of the star in the color-magnitude diagram compared to the locus of dSph members, quantified by CMD-based photometric metallicity estimates \citep[\fehp, e.g.,][]{gilbert2009gss,gilbert2012}. We update the halo star classifications in this work using more recent estimates of the mean velocity 
and velocity dispersion 
\citep[$\langle v \rangle$ and $\sigma_v$, respectively;][]{tollerud2012} and the half-light radius \citep{martin2016} for each dSph.  Stars are considered M31 halo stars if they are securely classified as RGB stars by the diagnostic method of \citet{gilbert2006}, and are more than 4$\sigma_v$ removed from the $\langle v \rangle$ of the dSph or lie more than 4 half-light radii in projected distance from the center of the dSph.   We note that the M31 dSph velocity dispersion measurements by \citet{tollerud2012} did not formally reject velocity outliers.   
In some cases, this resulted in a larger $\sigma_v$ than found in previous studies---or more recently, by \citet{kirby2020}. The \citet{tollerud2012} $\sigma_v$ estimates thus provide a conservative choice for halo star selection.

The only exception to the criteria listed above is applied in the And~I field.  This field is located on M31's Giant Stellar Stream (GSS) and has been shown to include GSS substructure at velocities similar to those of the And~I dSph, although the GSS population is comprised of significantly redder stars \citep{gilbert2009gss}.  This population can be seen in \fehp\ vs.\ heliocentric velocity space at \fehp\,$\gtrsim -1.0$.  Following \citet{gilbert2009gss}, RGB stars in the And~I field with \fehp\,$>$\,\andIfehphotselection\ were classified as M31 halo stars, rather than dSph members, regardless of their velocity or projected distance from the center of And~I.

The above selection criteria make no distinction between halo stars recently accreted into M31's stellar halo and those that were accreted at earlier times.  In particular, the selection criteria explicitly include both stars in known tidal debris features (in the And~I field) and potential extratidal stars that were recently stripped from the dSph in which they were born, in addition to stars in the relatively ``smooth'' halo of M31.  If a significant fraction of the M31 halo stars are in fact extratidal stars recently stripped from a surviving M31 dSph, and if M31’s outer stellar halo was built from dwarf galaxies with masses or star formation histories that are significantly different from those of these existing dwarf satellites, the properties of the current outer halo sample may not be representative of M31’s outer halo as a whole. 

\subsection{Measurements of \feh\ and \afe} \label{sec:abund_measurements}

Measurements of \feh\ and \afe\ were made using the methodology developed by \citet{kirby2008a,kirby2010}.  Each observed stellar spectrum is compared to a grid of synthetic spectra \citep{kirby2008a, kirby2010, kirby2011d} in order to determine the best-fitting values of effective temperature (\teff), \feh, and \afe\@.  This technique enables elemental abundances to be measured from relatively low-S/N spectra by leveraging all the abundance information in the spectrum simultaneously, including lines that are weak and/or blended in low- to moderate-resolution spectra.  

An initial continuum normalization via polynomial fitting was performed on each  wavelength-calibrated, sky-subtracted, one-dimensional spectrum after shifting it to the rest frame.  The surface gravity (log $g$) was held fixed based on a comparison of the star's position in the color-magnitude diagram with stellar evolutionary models.  Although the halo stars shown here could be at a range of line-of-sight distances corresponding to the depth of the M31 halo along a given sightline, the parameters of interest have been shown to be insensitive to the assumed distance modulus over reasonable ranges of line-of-sight distance for M31's halo \citep{vargas2014ApJL}. 

The determination of the abundances was an iterative process. We first fit for initial estimates of \teff\ and \feh, and then held these values constant while computing the bulk \afe\ abundance (from spectral regions sensitive to Mg, Si, Ca, and Ti). We next refined the continuum normalization by dividing the observed spectrum by the best-fit synthetic spectrum. These steps were then iterated until convergence of the continuum normalization, as well as the \teff, \feh\, and \afe\ values, was achieved.  Random uncertainties were estimated from the diagonal elements of the covariance matrix, and added in quadrature to the systematic uncertainty in \feh\ and \afe\ ($\sigma_{\rm sys} = 0.101$ and 0.084, respectively, determined as described by \citealt{kirby2010} and updated after modifying the abundance code, as described by \citealt{kirby2018}).

Further details of the abundance measurements of the deep and shallow observations are provided by \citet{kirby2020} and \citet{wojno2020}, respectively.  \citet{wojno2020} use the Python implementation \citep{escala2019} of the abundance measurement routine described by \citet{kirby2008a}. Comparisons of the measurements between the fitting codes used by \citet{kirby2020} and \citet{wojno2020} show that they return results consistent within the estimated uncertainties \citep{wojno2020}. 
  
In this analysis, we consider only measurements passing the following quality criteria.  In addition to convergence, well-constrained minima in the \teff\ and \feh\ $\chi^2$ contours are required for \feh\ measurements, and additionally in the \afe\ $\chi^2$ contours for \afe\ measurements. We regard the parameter as well-constrained if the $\chi^2$ contours extend smoothly from the minimum value to at least the $\pm1\sigma$ uncertainty thresholds; by definition, this excludes measurements with a minimum $\chi^2$ at the edge of the spectral grid. In addition, only \feh\ and \afe\ measurements with uncertainties $\le 0.4$~dex are included.  Finally, any stars showing TiO absorption features (at $\lambda\lambda \sim 7050$\,--\,7250\,\AA) are removed from the sample, as the fidelity of abundance measurements made with the current library of synthetic spectra (which do not include TiO features) has not been validated. The systematic effect on the sample of removing stars with TiO absorption is discussed in Section \ref{sec:cmdfeh}.  

For the remainder of this paper, \feh\ measurements derived using the spectral synthesis technique will be referred to as \fehsynth, to distinguish them from \feh\ estimates derived via other methods. 

The results of applying the M31 halo star sample selection criteria in conjunction with the abundance quality criteria are illustrated in Figures~\ref{fig:sampleselection1} and \ref{fig:sampleselection2}.  The selection criteria result in a total of \nouterhalofehours\ stars with \fehsynth\ measurements, in fields ranging from 43 to 165~kpc in projected distance from M31's center.  Of these, \nouterhaloafeours\ also have \afe\ measurements passing all quality criteria.  Table 1 %~\ref{table_abund} 
lists the abundance measurements for each star in the final sample.

\section{The \feh\ and \afe\ Distribution of Stars in M31's Outer Halo}\label{sec:halo_abundances}

\defcitealias{vargas2014ApJL}{V14}

\begin{figure}
\includegraphics[width=0.5\textwidth]{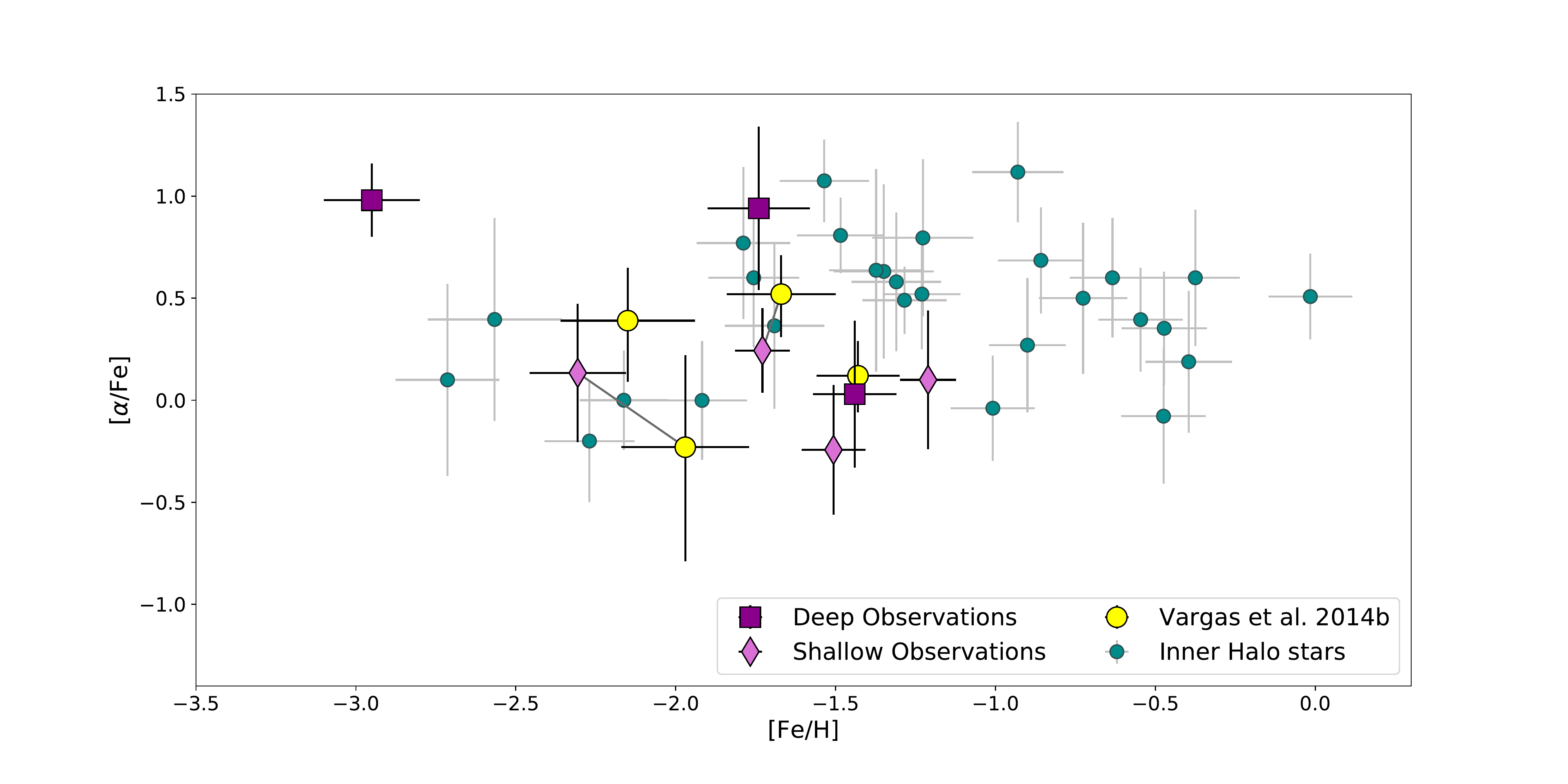}
\caption{The distribution of \afe\ vs.\ \fehsynth\ for the M31 halo sample (Section~\ref{sec:halo_abundances}).  Our M31 outer halo stars are shown as purple points; these \nouterhaloafeours\ measurements double the existing sample of four stars in M31's outer halo with \afe\ measurements \citep[yellow points;][]{vargas2014ApJL}.  Two of the stars with measurements in our sample were also in the M31 outer halo sample of \citeauthor{vargas2014ApJL}, and these are shown as connected points. In the remainder of the paper, we use our \fehsynth\ and \afe\ measurements for these two stars, and adjust the \fehsynth\ of the other two \citeauthor{vargas2014ApJL} stars by $-0.3$~dex, which is the mean systematic offset found between our measurements and those of \citealt{vargas2014} \citep[Section~\ref{sec:halo_abundances};][]{kirby2020}.  Also shown are measurements of stars in five inner halo fields that have velocities consistent with belonging to the dynamically hot component of M31's halo, rather than tidal debris features or M31's disk \citep{escala2019,escala2020,gilbert2019}.    
The outer halo stars span a similar range of \afe\ as the inner halo stars.  While the mean \afe\ of the outer halo sample is lower than that of the inner halo sample, the means are formally consistent within one sigma (Section~\ref{sec:halo_abundances}).   
}
\label{fig:m31halo_afe_feh}
\end{figure}

Figure~\ref{fig:m31halo_afe_feh} displays the \fehsynth\ and \afe\ measurements of the \nouterhaloafeours\ M31 outer halo stars passing all sample selection (Section~\ref{sec:halo_selection}) and measurement quality (Section~\ref{sec:abund_measurements})
criteria, as well as \fehsynth\ and \afe\ measurements of M31 halo stars from the literature.    

Of the four M31 outer halo stars with previous \afe\ and \fehsynth\ measurements \citep[hereafter V14]{vargas2014ApJL}, two are drawn from fields analyzed here\footnote{The other two \citeauthor{vargas2014ApJL} stars were observed in the And~II and And~XIII dSph fields.}.  Our outer halo sample includes measurements for these two \citetalias{vargas2014ApJL} stars. Figure~\ref{fig:m31halo_afe_feh} shows both our and \citeauthor{vargas2014ApJL}'s measurements for these stars, with each pair of duplicate measurements connected by a line.
The discrepancies in \fehsynth\ between our measurements and the \citetalias{vargas2014ApJL} measurements are fully consistent with a mean systematic offset of 0.3~dex (weighted by the inverse square of the quadrature sum of the uncertainties), with standard deviation of 0.32~dex, between our \fehsynth\ measurements and those of \citet{vargas2014}, with \citeauthor{vargas2014} being on average higher.  The mean \fehsynth\ offset was measured from an overlapping sample of M31 dSph stars, measured via the same abundance measurement pipelines as used for the outer halo samples \citep[Appendix A;][]{kirby2020}. \citeauthor{kirby2020} found no systematic offset between their \afe\ measurements and those of \citet{vargas2014}. The standard deviation of the difference in \afe\ measurements between the two samples was found to be 0.37 dex, and the measurement uncertainties were found to fully account for the scatter in \afe. The two halo stars in common between \citetalias{vargas2014ApJL} and our work, shown in Figure~\ref{fig:m31halo_afe_feh}, are consistent with the results of \citet{kirby2020}.  After accounting for the systematic offset in \fehsynth\ found between the two pipelines, the \fehsynth\ and \afe\ measurements from the two pipelines are consistent within the $\sim 1\sigma$
measurement uncertainties.

\begin{figure*}
\includegraphics[width=\textwidth]{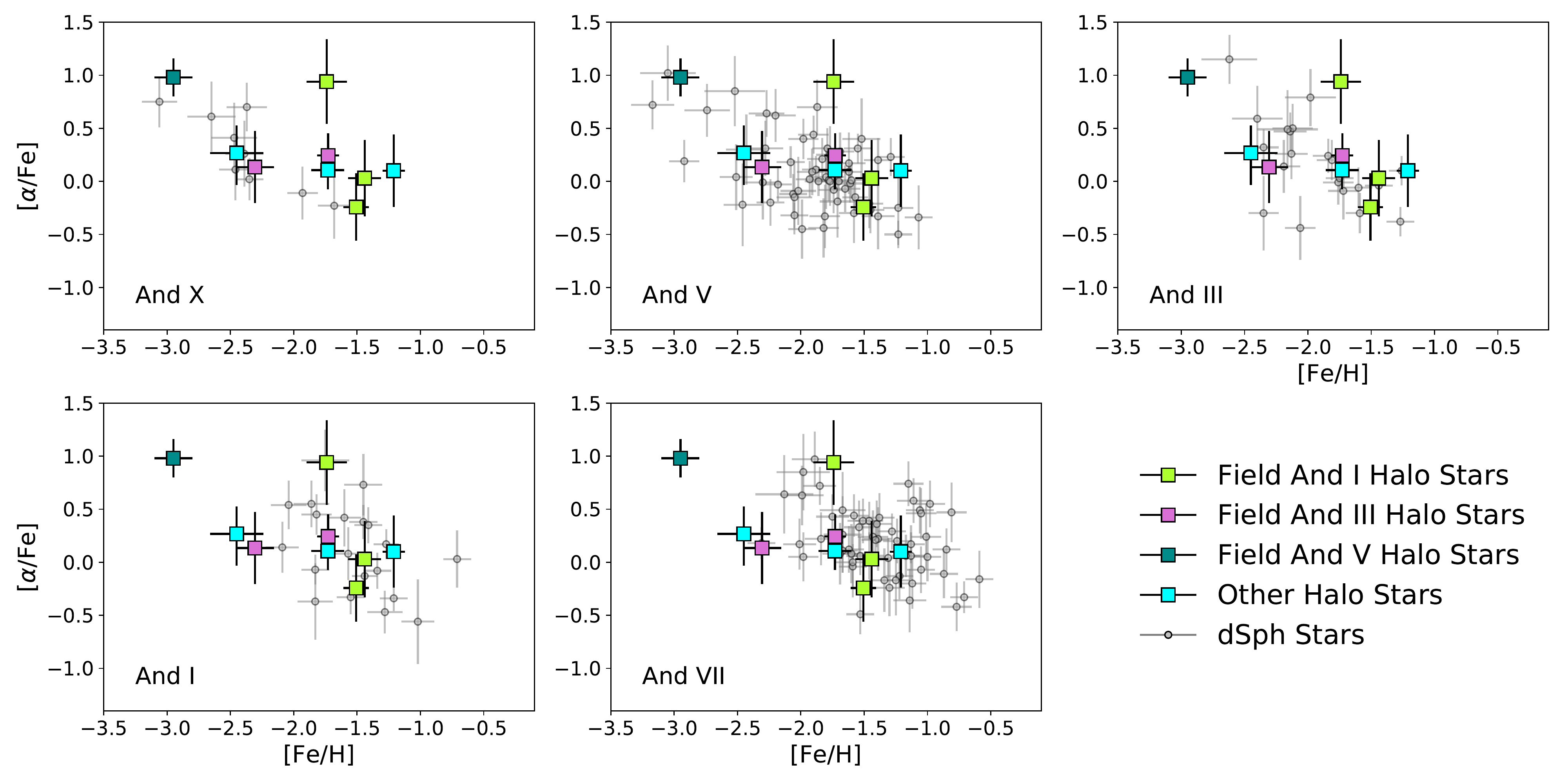}
\caption{Comparison of \afe\ vs.\ \fehsynth\ for M31 dSphs \citep[small, grey symbols][]{kirby2020} against the halo members (large colored symbols) presented in this work.  
The outer halo abundance measurements are fully consistent with the distribution of \afe\ vs. \fehsynth\ abundances of stars in M31 dSphs (Section~\ref{sec:halo_abundances}).   
}
\label{fig:m31halo_vs_dsph_afe}
\end{figure*}

\begin{figure*}
\includegraphics[width=\textwidth]{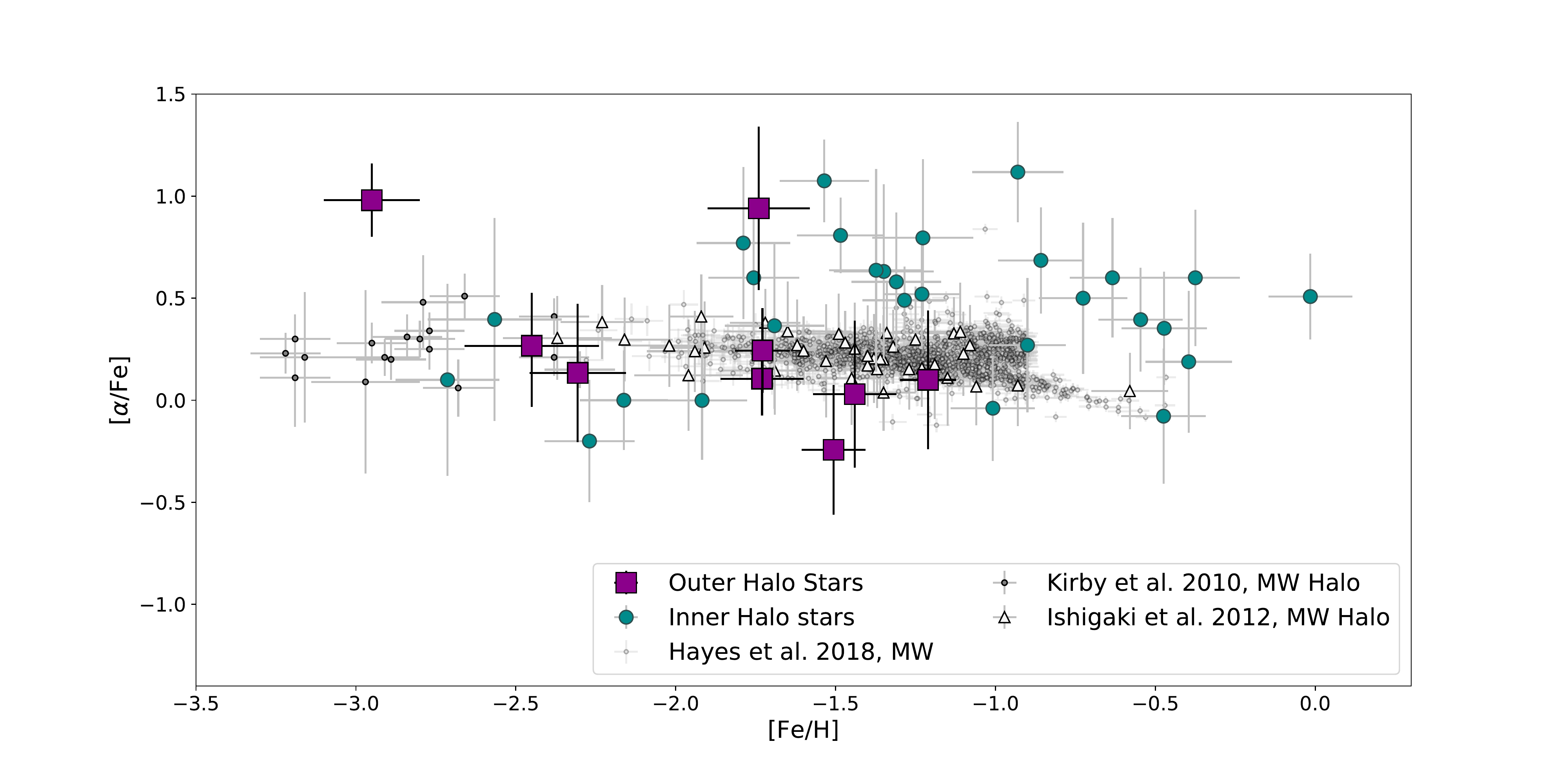}
\caption{Distribution of \afe\ vs.\ \fehsynth\ for stars in M31's inner and outer halo (as described in Figure~\ref{fig:m31halo_vs_dsph_afe}), compared to several samples of MW halo star abundances.  The \citet{hayes2018} measurements are the bulk \afe\ measurements reported by APOGEE's ASPCAP pipeline \citep{garciaperez2016} from the SDSS DR13 release \citep{albareti2017}, and exclude stars likely to be associated with the MW's thin and thick disks. The \citet{ishigaki2012} measurements have been converted to a bulk \afe\ estimate using equal weighting of the individual [Mg/Fe], [Si/Fe], [Ca/Fe], and [Ti/Fe] measurements.  The \citet{kirby2010} MW measurements were obtained using the same spectral synthesis technique used for the M31 measurements.  
On average, the abundance distribution of M31 outer halo stars appears to be more consistent with that of MW halo stars than with the abundance distribution of M31's kinematically hot inner halo  (Section~\ref{sec:halo_abundances}).
}
\label{fig:m31halo_vs_mw}
\end{figure*}

For the remainder of the paper, we use our measurements for the two stars that overlap the \citetalias{vargas2014ApJL} outer halo sample, and incorporate the remaining two \citetalias{vargas2014ApJL} stars into our analysis. To place the \citetalias{vargas2014ApJL} measurements on a scale consistent with our M31 halo star measurements, we make the following adjustments to the published \feh\ and \afe\ values. We adjust the \feh\ values by the mean systematic offset of $-0.3$~dex discussed above.  In Figures~\ref{fig:m31halo_vs_dsph_afe} and \ref{fig:m31halo_vs_mw}, we also adjust the \citetalias{vargas2014ApJL} \afe\ values  to remove an empirical correction applied by \citetalias{vargas2014ApJL} with the goal of better representing an unweighted average of [Mg/Fe], [Si/Fe], [Ca/Fe], and [Ti/Fe].  The magnitude of the applied correction was dependent on \feh. Removing the correction applied by \citetalias{vargas2014ApJL} lowers the \afe\ values of the four \citetalias{vargas2014ApJL} stars by $\lesssim 0.1$~dex from the values shown in Figure~\ref{fig:m31halo_afe_feh}.

The sample of outer halo stars with \afe\ measurements represents the relatively smooth outer halo of M31, as it does not contain any stars likely to be associated with known tidal debris features. The only field analyzed in this contribution in which tidal debris features have been previously identified is the And~I field \citep{gilbert2009gss}, which has two previously identified tidal debris features: one at $-383.3\pm 7.6$~\kms, with a velocity dispersion ($\sigma_v$) of $32.4^{+8.7}_{-6.9}$~\kms, which is the GSS, and a faint feature at $-288.7^{+5.6}_{-6.1}$~\kms, with a velocity dispersion of $11.0^{+9.6}_{-4.5}$~\kms\ \citep{gilbert2018}.  Both features are comprised of stars with red colors, and are thus expected to be quite metal-rich.  Of the three stars in the And~I field with \afe\ estimates, one of the stars is very far removed in velocity space from either of the tidal debris features in the And~I field. The other two stars, while each less than $2\sigma_v$ removed from the mean velocity of one of the tidal debris features in the And I field, land in a very different position in the color-magnitude diagram than that of the stars associated with the tidal debris in And~I.  The stars with \afe\ measurements are significantly bluer, with metallicity estimates (both \fehp\ and \fehsynth) significantly lower than that expected for stars in the tidal debris features, based on \fehp\ estimates as well as \fehsynth\ measurements of the GSS in other fields \citep{gilbert2019,escala2020}.

The inner halo sample in Figure~\ref{fig:m31halo_afe_feh} is drawn from five fields in M31's inner halo, all within 30~kpc in projected distance from M31's center and observed with Keck/DEIMOS \citep[Figure~1;][]{escala2019,escala2020,gilbert2019}.  These fields contain numerous tidal debris features, including the GSS.  The majority of the measurements in the inner halo \citep[the four fields published by][]{escala2019,escala2020} were made from spectra obtained with the lower-resolution 600 line mm$^{-1}$ (600ZD) grating ($R\sim 2500$, $\lambda\lambda\sim 4500-9100$\AA).  The \citet{gilbert2019} inner halo field measurements were made from spectra obtained with the same instrumental setup as used for our outer halo measurements (namely, with the 1200 line mm$^{-1}$ grating).  \citet{escala2020} compare \feh\ and \afe\ measurements from a sample of stars observed with both the 1200G and 600ZD gratings; they find that both the \feh\ and \afe\ measurements are broadly consistent, with no statistically significant offsets between the two measurements.

As the outer halo stars with \afe\ measurements are unlikely to be associated with known tidal debris features, the inner halo sample shown in Figure~\ref{fig:m31halo_afe_feh} includes only stars with a high probability of belonging to the underlying, dynamically hot halo component ($P_{\rm halo}\geq 0.8$). The value of $P_{\rm halo}$ is computed using the best-fit model of the velocity distribution of M31 stars in the field \citep{gilbert2019,escala2020}. The outer halo points shown in Figure~\ref{fig:m31halo_afe_feh} span a range of projected distance from M31's center of \minrprojafeouterhalo\ to \maxrprojafeouterhalo~kpc, while the inner halo stars span a range of \minrprojinnerhalo\ to \maxrprojinnerhalo~kpc.  Four of the five inner halo fields are drawn from on or near M31's southeast minor axis or the Giant Stellar Stream (to the southwest of the minor axis fields).  The exception is the disk field, which is located near M31's northeast major axis, at 26~kpc.  

The inner and outer halo star samples span a similar range of \afe\ values.  To compare the mean (or median) \afe\ of the inner and outer halo samples, we account for the sampling uncertainty due to limited sample sizes using bootstrap resampling of the data. For each draw, $d$, a mean (or median), \meanafe$_{d}$, was calculated using a random sampling (with replacement) of the available measurements. For each measurement $i$ to be used in a given draw, the \afe$_{i}$ value used in computing \meanafe$_{d}$ was drawn from a Gaussian distribution with mean and dispersion given by the \afe\ measurement and its associated uncertainty for star $i$.  We report as \meanafe\ the mean of the \meanafe$_{d}$ values, with the uncertainty given by the standard deviation of the \meanafe$_{d}$ values, calculated using \nbootstrapdraws\ draws.

Using this process, we find mean \afe\ values of \meanafe\,$=$\,\meanafeinner~dex for the kinematically hot inner halo sample\footnote{This value is fully consistent with the \meanafe\ computed without imposing a $P_{\rm halo}$ cut on the inner halo sample.  If the mean \afe\ is computed from the full inner halo star sample, weighting each star by $P_{\rm halo}$, as well as the uncertainty in its \afe\ measurement, \meanafe\,$=$\,\weightedmeanafeinner~dex.}, and \meanafeouter\ for the outer halo points (\meanafepastseventy\ if only stars beyond 70~kpc in projected distance from M31's center are considered).  Thus, while the outer halo stars are, on average, less $\alpha$-enhanced than the inner halo stars, the \meanafe\ of the current inner and outer M31 halo samples are formally consistent with each other.  This remains true when restricting the inner halo sample to stars within the same range of \fehsynth\ as observed in the outer halo sample: for inner halo stars with \fehsynth$<-1.1$,  \meanafe\,$=$\,\meanafeinnerfehrestricted~dex. 

Figure~\ref{fig:m31halo_vs_dsph_afe} shows the distribution of \afe\ vs.\ \fehsynth\ abundances of the M31 outer halo stars compared to the abundance distributions of five M31 dSphs based on deep spectroscopic observations \citep{kirby2020}.  The dSphs in Figure~\ref{fig:m31halo_vs_dsph_afe} span a range of estimated stellar mass of (0.12\,--\,16)$\times 10^6$\Msun, assuming the luminosity estimates computed by \citet{tollerud2012} and the stellar mass-to-light ratios compiled by \citet{woo2008}, as described in \citet{kirby2020}.  In contrast to the inner halo of M31 \citep{gilbert2019,escala2020}, the current M31 outer halo sample is fully consistent with the range of \fehsynth\ abundances, as well as the trend of \afe\ as a function of \fehsynth, measured in the largest of these five M31 dSphs (e.g., And~I and And~VII).  This may indicate that M31's outer halo was built from progenitors with stellar masses and star formation histories similar to those of And~I and And~VII, although a larger sample of M31 outer halo star abundances, including stars from fields not associated with M31 dSphs, is necessary to draw any firm and quantitative conclusions.    

Figure~\ref{fig:m31halo_vs_mw} compares the \afe\ and \feh\ abundances of stars in M31's dynamically hot inner and outer halo with measurements of stars in the MW. The \citet{hayes2018} low-metallicity MW sample is drawn from Apache Point Observatory Galactic Evolution Experiment \citep[APOGEE;][]{majewski2017} data ($R\sim 22,500$) presented in the Sloan Digital Sky Survey \citep[SDSS-III;][]{eisenstein2011} Data Release 13 (DR13; Albareti et al. 2017).  The \citeauthor{hayes2018}\ MW sample excludes stars likely associated with the MW's thin and thick disks, although \citeauthor{hayes2018}\ argue that a portion of the high \afe\ stars may have their origin in, or share a formation history with, the thick disk.
The \citet{kirby2010} MW halo star measurements were derived from observations obtained with Keck/DEIMOS, in the same configuration as for the M31 outer halo measurements and using an earlier version of the same abundance measurement routines.   
The \citet{ishigaki2012} measurements were derived from high-resolution spectra ($R\sim 50,000$ or $R\sim 90,000$) obtained with the High Dispersion Spectrograph on the Subaru Telescope, using an LTE abundance analysis to measure the abundances of individual $\alpha$-elements.  The \citet{ishigaki2012} sample consists of nearby stars (within 2~kpc of the Sun) with kinematics consistent with belonging to the stellar halo of the MW. 

Stars in the dynamically hot component of M31's inner halo appear to be, on average, significantly $\alpha$-enhanced compared to the MW halo stars, specifically for stars with $-1.8\lesssim$\,\fehsynth\,$\lesssim -1.0$.  The M31 inner halo sample also extends to  larger \feh\ values than does the MW halo sample of \citet{ishigaki2012}.  The five M31 inner halo stars with \fehsynth$\lesssim -1.8$ appear to have  $\alpha$-enhancement similar to that of the stars in the MW sample.  

In contrast, the \afe\ abundances of stars in M31's outer halo appear more broadly consistent with the abundances of MW halo stars.  The M31 outer halo stars span a range of \feh\ similar to that of the MW halo stars.  The M31 outer halo stars are also more consistent, on average, with the \afe\ abundances of the MW halo sample, with a smaller fraction of stars with high \afe\ measurements than found in M31's inner halo. This may indicate that the abundances of stars in M31's outer stellar halo are more similar to the abundances of stars in the MW's halo than to stars in M31's inner halo.  This would be consistent with other indications that the properties of M31's outer stellar halo are fairly similar to the MW's stellar halo, while M31's inner halo is markedly different, e.g., in terms of surface brightness profile and mean photometric metallicity \citep{irwin2005,gilbert2012,gilbert2014,ibata2014}.  However, a rigorous comparison of MW and M31 halo abundances will require a larger sample of M31 halo stars, especially in the outer halo.

\section{Comparison of Trends in Spectroscopic and CMD-based \feh\ Distributions}\label{sec:cmdfeh}
\defcitealias{gilbert2014}{G14}

\begin{figure}
\includegraphics[width=0.49\textwidth]{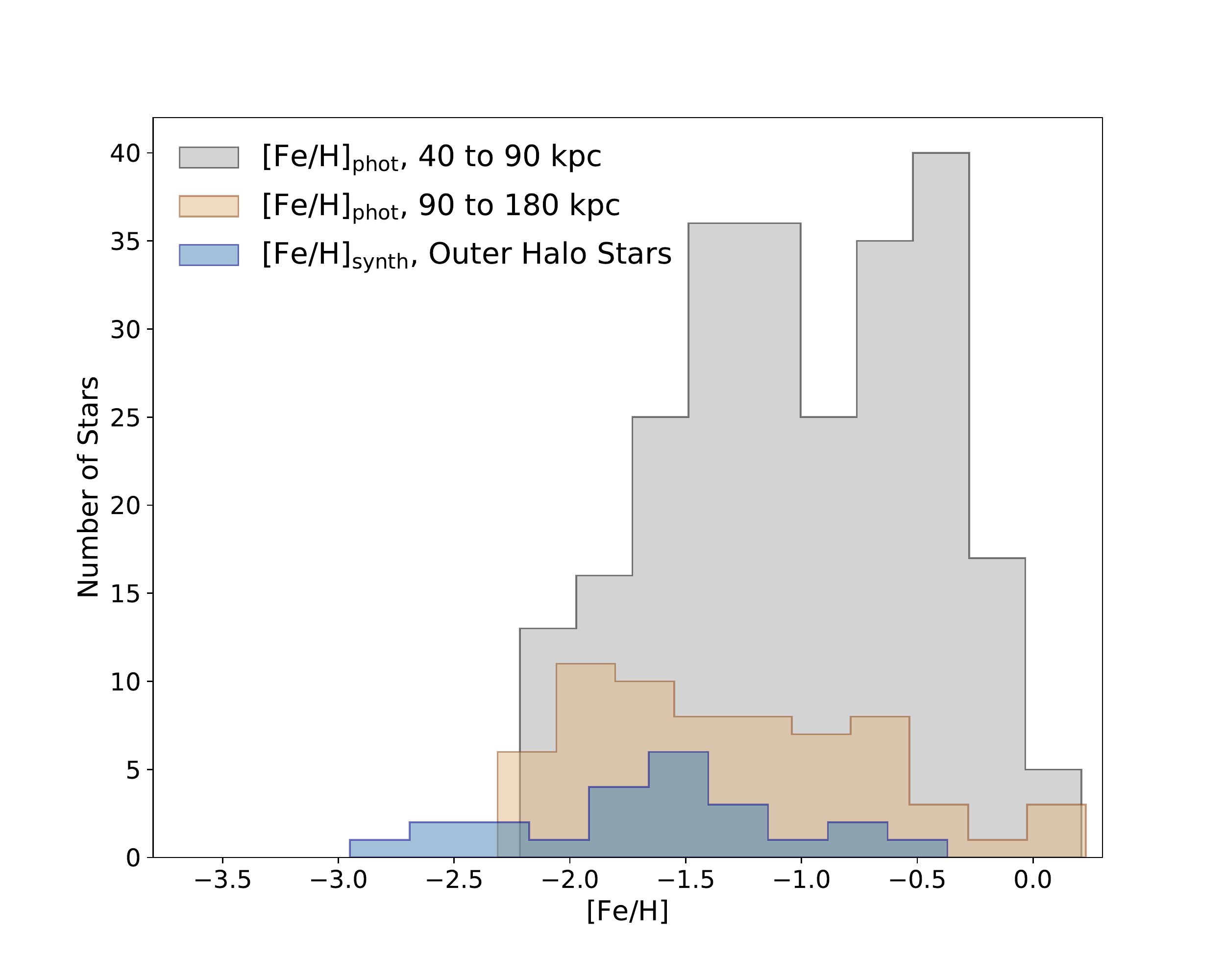}
\includegraphics[width=0.49\textwidth]{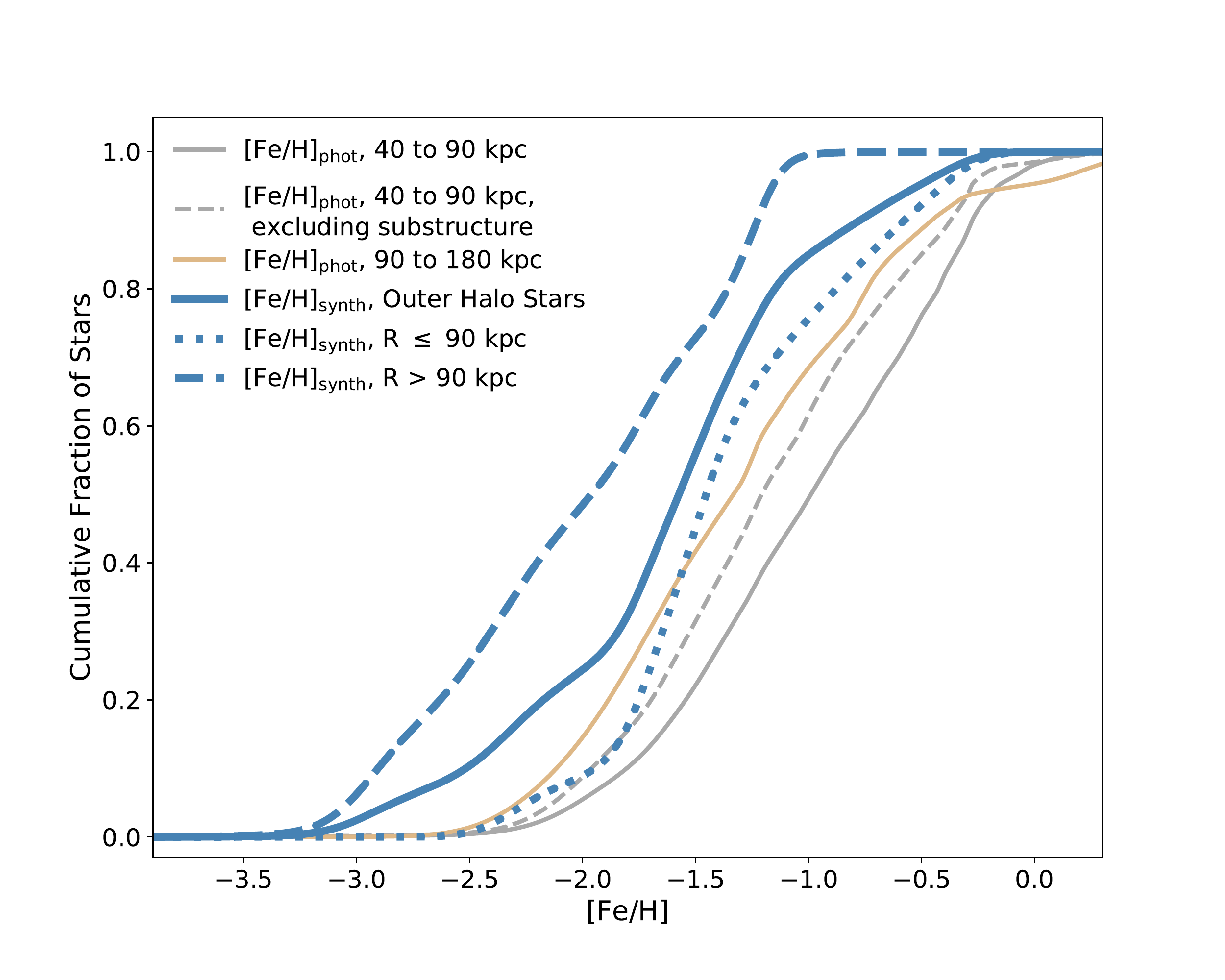}
\caption{Top: The M31 outer halo \fehsynth\ measurements (blue histogram)
compared to CMD-based \fehp\ measurements for all securely identified M31 halo stars in the SPLASH survey with projected distances from M31 that place them in the two outer radial bins analyzed by \citet{gilbert2014}.  These two radial bins span the full range of \rproj\ of the M31 outer halo stars with \fehsynth\ measurements.  The outer halo sample \fehsynth\ measurements span a large range of \feh, as is also seen in the \fehp\ sample.   
Bottom: Cumulative \feh\ distributions for each sample.  Also shown are the \fehsynth\ distributions of the outer halo stars in bins of \rproj\ matching those used by \citetalias{gilbert2014} (blue dashed and dotted curves).   The \fehsynth\ outer halo distributions appear to be, on average, more metal-poor than the \fehp\ distributions.
However, there are several sources of bias in the samples shown that preclude a simple inference based on these comparisons (Section~\ref{sec:cmdfeh}).
}
\label{fig:m31halo_feh_radialbins}
\end{figure}

As discussed in Section~\ref{sec:intro}, previous measurements of the metallicity of stars in M31's outer stellar halo have largely been based on \fehp\ estimates, the most recent being  \citet{gilbert2014} and \citet{ibata2014}.    
Such metallicity estimates are sensitive to the assumed (singular) age and $\alpha$-enrichment of the stellar population, and any change in the mean age or $\alpha$-enrichment of the stellar population as a function of radius will affect the accuracy of a metallicity gradient measured from \fehp\ abundances \citep[hereafter G14]{gilbert2014}.  For the \citetalias{gilbert2014} M31 halo sample, changing the assumed age or \afe\ used in computing \fehp\, or adopting alternative stellar evolution models\footnote{\citetalias{gilbert2014} computed \fehp\ using models from the Padova group \citep{girardi2002} and the Yale-Yonsei group \citep{demarque2004} in addition to the nominal \citet{vandenberg2006} isochrones.  We have also computed \fehp\ for the \citetalias{gilbert2014} halo sample using the PARSEC isochrones \citep{marigo2017}.} results in systematic offsets on the order of 0.1 to 0.2~dex in \fehp\ for most assumptions, with the magnitude of the offsets showing mild dependence on \feh\ \citepalias{gilbert2014}.     

\citetalias{gilbert2014} also estimated \feh\ from the EW of the \caii\ triplet absorption feature (\fehcat) via the empirical calibration by \citet{ho2012}, which uses the strongest two \caii\ triplet lines at 8542 and 8662\AA. While the \fehcat\ estimates did not require an assumed age or \afe\ abundance, the two absorption lines on which they were based are frequently affected by night sky lines at the line-of-sight velocities typical of M31 halo stars.  This, combined with the relatively low S/N of stars in the \citetalias{gilbert2014} sample, led to high uncertainties in the individual \fehcat\ estimates (on the order of 0.5~dex). Nevertheless, \citetalias{gilbert2014} found reasonable agreement in the means of the \fehp\ and \fehcat\ distributions as a function of \rproj.  While both the \fehp\ and \fehcat\ estimates depend on assuming a line-of-sight distance, assuming a star is $\pm 100$~kpc from M31's distance results in a shift in metallicity of the same order as the random uncertainties (or of the systematic uncertainties) for \fehp, and significantly less than the random uncertainties for \fehcat\ \citep{gilbert2014}.  Moreover, since M31's stellar halo surface brightness profile is highly concentrated, the magnitude of this effect will be significantly smaller for the vast majority of halo stars.        

Unlike \fehp, the \fehsynth\ measurements do not require assuming a single \afe\ for the stellar population, and are relatively insensitive to the assumed age, since log $g$, which is largely insensitive to stellar age, is the only stellar parameter that is held fixed based on the photometric measurements.  The \fehsynth\ measurements have also been shown to be insensitive to the assumed line-of-sight distance over reasonable ranges for M31's halo \citep{vargas2014ApJL}.
Moreover, with mean uncertainties of \meanfehuncertainties~dex, the \fehsynth\ measurements presented here are more precise than the \fehcat\ measurements presented by \citetalias{gilbert2014}.  They thus provide an important independent, direct estimate of \feh\ in M31's outer halo. 

Figure~\ref{fig:m31halo_feh_radialbins} compares the \fehsynth\ distribution of the outer halo star sample to the \fehp\ distributions of M31 halo stars in two large bins of projected distance from M31 (40\,--\,90~kpc and 90\,--\,180~kpc).  These two bins encompass the range of distances of the M31 outer halo stars with \fehsynth\ measurements (\minrprojfehouterhalo\,--\,\maxrprojfehouterhalo~kpc).  The \fehp\ values are those published by \citetalias{gilbert2014} and were calculated by comparing each star's position in the \ivi\ CMD to the \citet{vandenberg2006} stellar evolutionary models, assuming an age of 10~Gyr and solar \afe. The outer halo \fehsynth\ measurements span a large range of \feh\ values, although the 
distribution is on average more metal-poor than the \citetalias{gilbert2014} \fehp\ distributions.   
The cumulative distributions in the bottom panel of Figure~\ref{fig:m31halo_feh_radialbins} were constructed using a sum of Gaussians, with means equal to the \feh\ measurement and standard deviation equal to the uncertainty in the \feh\ measurement.

There are several subtleties to consider when comparing the distributions in Figure~\ref{fig:m31halo_feh_radialbins}.  The first is measurement bias in the distributions. The \fehp\ distributions have been truncated to \fehp$\geq -2.32$~dex, corresponding to the most metal-poor isochrone \citep{vandenberg2006}.  If the outer halo \fehsynth\ values are similarly truncated, the metal-poor end of the \fehsynth\ distribution is similar to the 90\,--\,180~kpc \fehp\ bin for \feh\,$\lesssim -1.5$~dex.  We also expect bias in the \fehsynth\ distribution at the metal-rich end. The innermost outer halo field (field And~I) has a significant population of securely identified M31 halo stars with \vio\,$\gtrsim 2$ (Figure~\ref{fig:sampleselection1}), where either the typically lower spectral S/N results in the spectral synthesis measurement failing to pass all quality criteria, or molecular TiO absorption bands are strong, necessitating removal of the star from the sample \citep{gilbert2019, escala2020}. Thus, we anticipate that the \fehsynth\ distribution in Figure~\ref{fig:m31halo_feh_radialbins}, for stars with \rproj\,$<90$~kpc, includes a significant bias against metal-rich stars, which we discuss further below.  

The second factor affecting the comparisons in Figure~\ref{fig:m31halo_feh_radialbins} is the radial distribution of the stars included in each \feh\ distribution.  
Because M31's halo has been observed to contain a significant metallicity gradient \citep{kalirai2006halo,koch2008,gilbert2014,ibata2014}, \feh\ distributions covering broad ranges of projected distance from M31, such as those in Figure~\ref{fig:m31halo_feh_radialbins}, will be sensitive to the distribution of \rproj\ of the stars in the bin.

\begin{figure*}
\includegraphics[width=\textwidth]{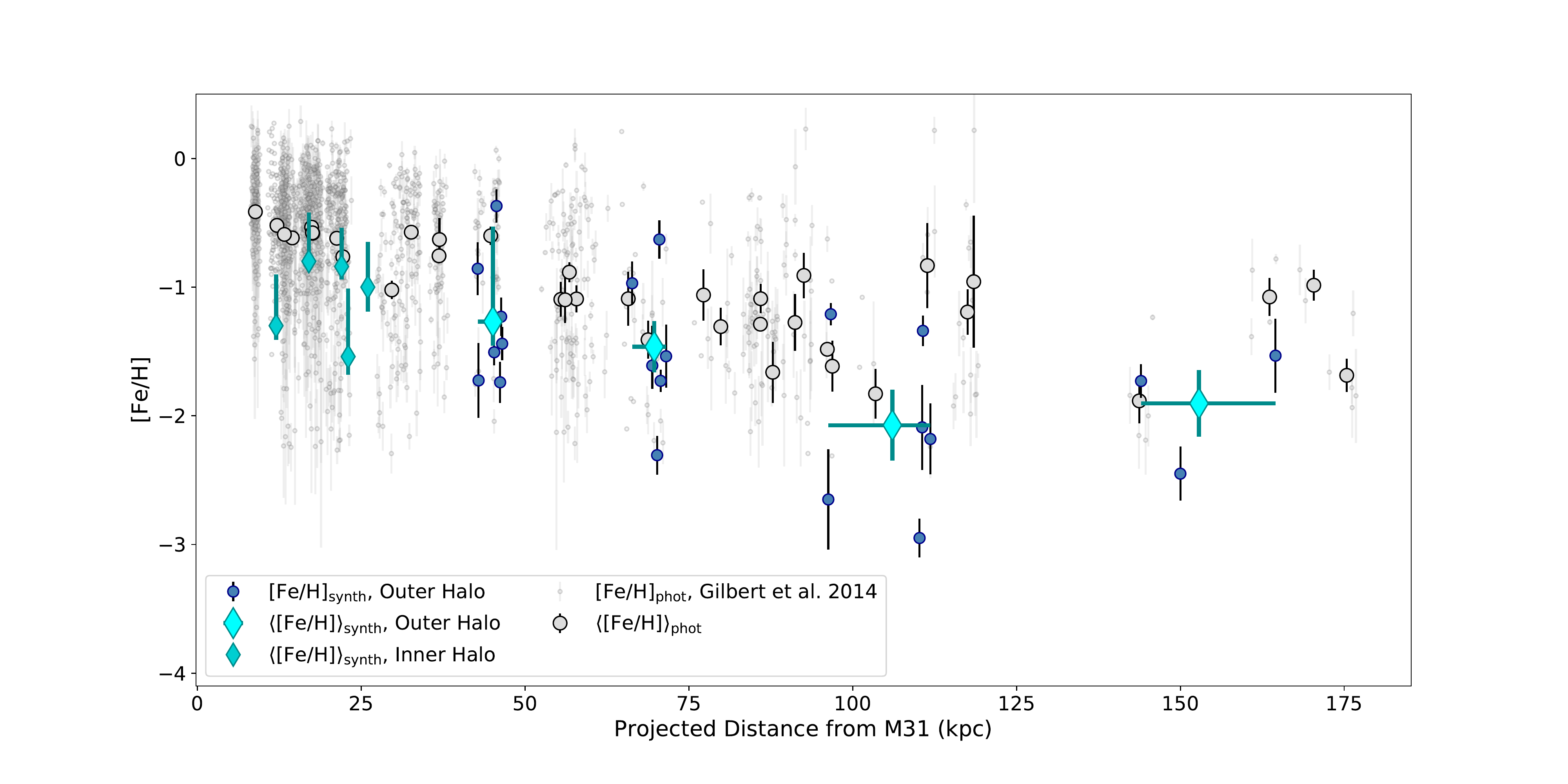}
\caption{M31 outer halo spectral synthesis \fehsynth\ measurements (small blue points)
compared to the \citet{gilbert2014} CMD-based \fehp\ measurements for all securely identified M31 halo stars in the SPLASH survey (small gray points).  
Large cyan points show the mean of the \fehsynth\ measurements (\meanfehsynth) as a function of projected distance from M31, shown at the mean \rproj\ of the stars; the error bars on the abscissa denote the range of \rproj\ of the stars contributing to \meanfehsynth.  The \meanfehsynth\ values for fields at \rproj\,$\lesssim$\,\maxrprojinnerhalo~kpc are computed from the measurements by \citet{escala2020} and \citet{gilbert2019}.  Large gray points are the mean \fehp\ of all M31 RGB stars in each \citet{gilbert2014} field.  
At a given \rproj, the current outer halo \fehsynth\ measurements show reasonable agreement with the mean \fehp\ and range of \fehp\ values.  While the \meanfehsynth\ values appear systematically lower than the \fehp\ sample as a whole, there are currently large uncertainties in the magnitude of the systematic bias against recovery of red (likely metal-rich) stars in fields at \rproj$<50$~kpc, as indicated by the asymmetric error bars on \fehsynth\ (Section~\ref{sec:cmdfeh}).
Nevertheless, the \fehsynth\ abundances are consistent with previous findings that M31's halo becomes increasingly more metal-poor on average with increasing \rproj, out to \rproj\,$\sim 100$~kpc.
\label{fig:m31halo_feh_vs_rad}
}
\end{figure*}

Thus, it is instructive to look at the \feh\ abundances as a function of projected distance from M31's center. 
Figure~\ref{fig:m31halo_feh_vs_rad} shows the \fehp\ \citepalias{gilbert2014} and the \fehsynth\ abundances of all M31 halo stars (including substructure) as a function of projected distance from M31, as well as the mean \fehp\ for each \citetalias{gilbert2014} spectroscopic field and means of the \fehsynth\ sample.  
To properly account for the small \fehsynth\ samples, especially in the outer halo, the mean \fehsynth\ measurements shown in Figure~\ref{fig:m31halo_feh_vs_rad} were computed in bins of projected distance using bootstrap resampling of the data points in each bin, as described in Section~\ref{sec:halo_abundances}.  

The mean \fehsynth\ in M31's outer halo (\rproj\,$\geq$\,\minrprojfehouterhalo~kpc for our sample) becomes increasingly metal-poor on average with increasing \rproj, out to \rproj$\sim 100$~kpc, following the trend previously observed in \fehp.  The mean \fehsynth\ of the outer halo fields are consistent with the mean \fehp\ value of the closest spectroscopic field in \rproj, with the exception of field And~I at \rproj\,$\sim$\,\rprojandone~kpc, discussed in more detail below.  Given the limited sampling of the distribution, the spread of \fehsynth\ measurements in our outer halo fields (\rproj\,$\geq$\minrprojfehouterhalo~kpc) is also largely consistent with the spread of \fehp\ estimates at a given \rproj.  However,  
in aggregate, the outer halo \meanfehsynth\ values appear to be systematically lower than the mean \fehp\ trend line. 

Also shown in Figure~\ref{fig:m31halo_feh_vs_rad} are the mean \fehsynth\ values for the five inner halo fields \citep[\rproj$<30$~kpc;][]{gilbert2019,escala2020}.   
Three of the inner halo fields (at \rprojH, \rprojkmggss, and \rprojS~kpc) include known substructure in M31, primarily related to M31's Giant Stellar Stream; the abundance distributions as a function of component are discussed in \citet{gilbert2019} and \citet{escala2020}. 
The inner halo field at \rprojdisk~kpc is located in M31's outer disk; for this field, the mean \fehsynth\ shown is for M31's halo component, not for the field as a whole. 

In the inner fields, as many as 30--40\% of the stars with otherwise successful \fehsynth\ measurements were not included in the final sample due to the presence of strong TiO absorption in the stellar spectrum, which is not modeled in the synthetic spectra.  The majority of these stars are red [$(g'-i')_0$ or \vio\,$\gtrsim 2$~mag] and are expected to be metal-rich.  Moreover, there is an additional (although smaller) bias due to a higher rate of successful \fehsynth\ measurements for brighter, more metal-poor RGB stars.  The asymmetric error bars on the inner field values indicate the lower uncertainty in the mean, and the upper uncertainty in the mean plus an estimate of the potential systematic bias against recovering \fehsynth\ for metal-rich stars. 
The systematic bias estimates are based on \fehp\ estimates, computed using the PARSEC isochrones \citep{marigo2017}, which include the effects of TiO absorption, for stars with failed or unreliable \fehsynth, as described in \citet{escala2020}.

The mean \fehsynth\ of three of the inner fields are fully consistent with the mean \fehp\ measurements of previous spectroscopic fields at comparable \rproj.  However, two inner halo fields (at \rprojH~kpc and \rprojiehalo~kpc on the minor axis) have significantly lower mean \fehsynth\ than the corresponding mean \fehp\ measurement of either the original, shallow spectroscopic field, or spectroscopic fields at comparable \rproj. 
One of these fields (\rprojiehalo~kpc) contains no identified tidal debris features, while the other (\rprojH~kpc) is located on the SE shelf feature thought to be an extension of the Giant Stellar Stream \citep{fardal2007,gilbert2007}.  

It is possible that real and significant variation in the mean \feh\ between closely spaced fields in M31's inner halo is being revealed by the spectral synthesis measurements. 
However, the systematic biases affecting the measurement of the mean \fehsynth\  
make it difficult to judge the significance of the field-by-field \fehsynth\ differences in the inner halo.  

The bias against measuring \fehsynth\ for metal-rich stars is less severe in M31's outer halo.  Only the outer halo field at \rprojandone~kpc (field And~I) is expected to suffer significantly from a bias against the measurement of red stars that are expected to be metal-rich (based on the colors and magnitudes of the stars; see Figures~\ref{fig:sampleselection1} and \ref{fig:sampleselection2}). Of the outer halo fields, field And~I is also the one that has the largest discrepancy between the mean \fehp\ and \fehsynth\ values.   Only two stars were removed from the outer halo \fehsynth\ sample due to the presence of TiO in the spectrum ($<10$\% of the total outer halo \fehsynth\ sample), but both were from the And~I field, resulting in the removal of 22\% of the stars with successful \fehsynth\ measurements in this field. 
The And~I field 
is also the only outer halo field that contains a significant population of likely M31 RGB stars with \vio\,$\gtrsim 2$, none of which resulted in \fehsynth\ measurements passing all quality criteria.  The systematic bias in our mean \fehsynth\ measurement in this field is estimated to be as large as \andIphotbias~dex. 

As discussed in Section~\ref{sec:halo_selection}, the And~I field is spatially coincident with M31's GSS\@.
The GSS comprises $56\pm13$\% of the halo population in this field \citep{gilbert2018}. However, only one of the \nAndIhalofeh\ \fehsynth\ measurements (14\%) in this field is highly probable to be associated with the GSS, falling within one $\sigma_v$ of the mean velocity of the GSS ($\langle v \rangle = -383.3$~\kms, dispersion $\sigma_v$ = 32.4~\kms; see Table 4 of \citet{gilbert2018}), while an additional two stars are both within $2\sigma_v$ of the mean velocity of the GSS and in a CMD region compatible with that of the GSS \citep[Figure~\ref{fig:sampleselection1}, Section~\ref{sec:halo_abundances};][]{gilbert2009gss, gilbert2018}. 
If stars potentially associated with substructure in this field are removed, the mean \fehp\ for this field is reduced from $-0.60\pm0.07$ to $-0.86\pm0.25$~dex, while the mean \fehsynth\ is reduced from \meanfehandIall\ to \meanfehandIexcltwosigma~dex.

\begin{figure}
\includegraphics[width=0.5\textwidth]{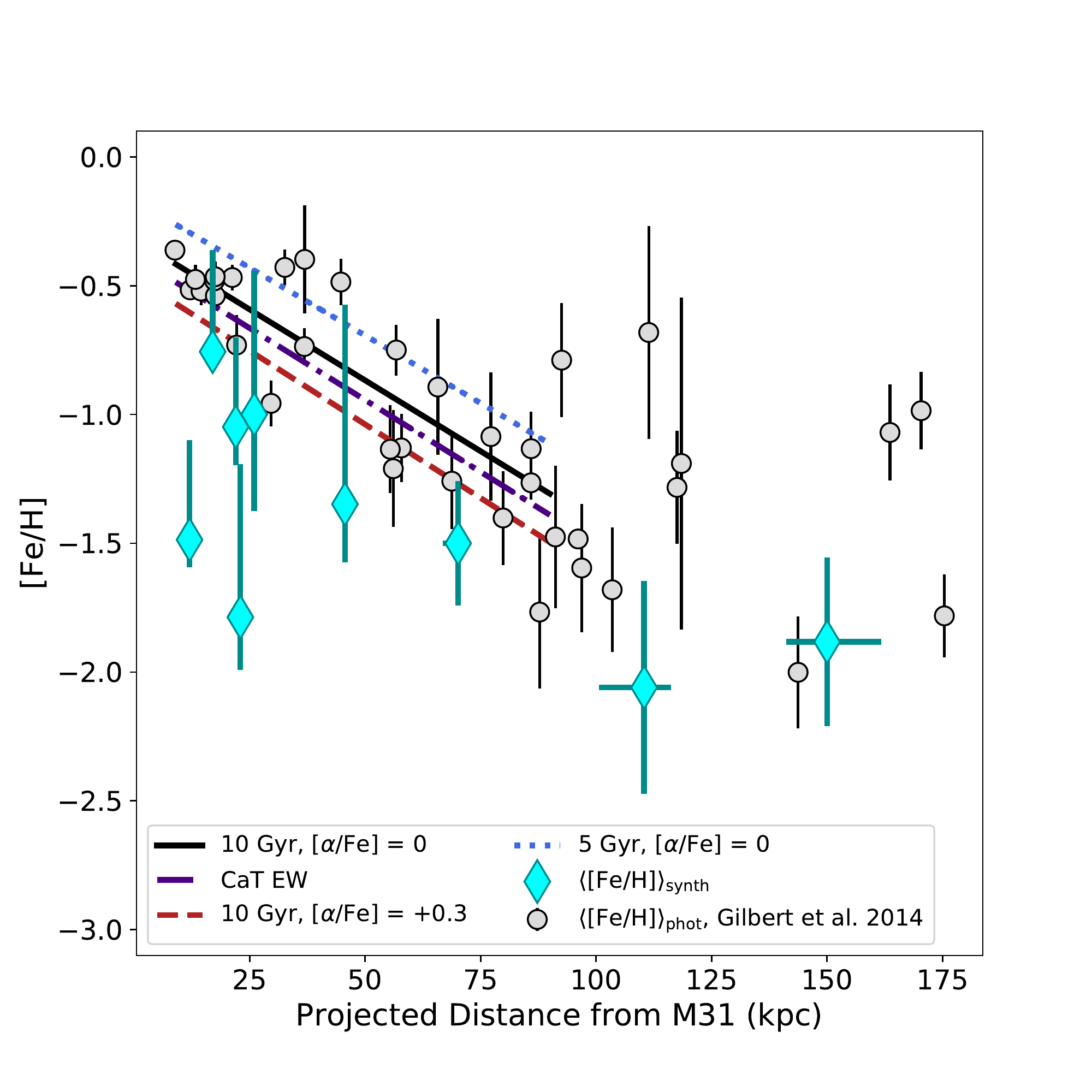}
\caption{
Comparison of the \fehsynth\ measurements with the metallicity gradient of M31's stellar halo as measured by \citet{gilbert2014}.  The nominal \citeauthor{gilbert2014} gradient was measured assuming 10~Gyr, \afe\,$=0$ isochrones (black line); the impact on the metallicity gradient of  adopting alternative assumptions when estimating \fehp is also shown, as well as the gradient measured using \fehcat\ estimates derived from the EW of the \caii\ triplet absorption feature for a subset of stars with relatively high S/N spectra \citep{gilbert2014}.  The large grey points show the median \fehp, assuming 10~Gyr, \afe\,$=0$ isochrones, of all M31 RGB stars in each \citet{gilbert2014} field. 
Cyan points show the median \fehsynth\ for our inner and outer halo fields (\citealt{gilbert2019}, \citealt{escala2020}, and this work).  The median \fehsynth\ measurements are most consistent with the gradient computed assuming $\alpha$-enhanced isochrones, as would be expected given the mean $\alpha$-enhancement observed in the inner and outer halo stars (Section~\ref{sec:halo_abundances}).  
\label{fig:m31halo_feh_vs_rad_fits}
}
\end{figure}

Finally, we compare the \fehsynth\ outer halo measurements with the halo metallicity gradient measured by \citetalias{gilbert2014}, which has a magnitude of $\sim -0.01$~dex~kpc$^{-1}$ (with or without the inclusion of substructure).   Figure~\ref{fig:m31halo_feh_vs_rad_fits} compares the median \fehsynth\ values in the outer halo with the nominal metallicity gradient for M31's stellar halo (including substructure), as well as gradients measured using alternate assumptions \citepalias{gilbert2014}.  Also shown in Figure~\ref{fig:m31halo_feh_vs_rad_fits} are the median \fehsynth\ abundances for fields in M31's inner halo \citep{escala2019,escala2020,gilbert2019}.  All median \fehsynth\ measurements, including those in the inner halo fields, were computed using bootstrap resampling as described in Section~\ref{sec:halo_abundances}.
All M31 halo stars, whether or not they are associated with tidal debris features, are included in the calculation of the median \fehsynth, for consistency with the analysis by \citetalias{gilbert2014}.

The \citetalias{gilbert2014} metallicity gradients were measured using spectroscopic fields ranging from 9 to 90 kpc in projected distance from M31's center, and were computed using the median \fehp\ of all M31 halo stars in each field under varying assumptions of age and $\alpha$-enhancement, as well as using the median \fehcat.   
The effect of assuming a different age and $\alpha$-enhancement on the measured \fehp\ gradient is  clearly shown in Figure~\ref{fig:m31halo_feh_vs_rad_fits}. A gradient in mean age or $\alpha$-enhancement of the stellar halo population with \rproj\ could cause the true \feh\ gradient to be steeper or shallower than that measured assuming a single age and $\alpha$-enhancement at all radii.

Consistent with the comparison of the \fehp\ and \fehsynth\ measurements in Figures~\ref{fig:m31halo_feh_radialbins} and \ref{fig:m31halo_feh_vs_rad}, the median \fehsynth\ values are systematically lower than the nominal gradient.  They are most consistent with the gradient measured assuming $\alpha$-enhanced isochrones (10 Gyr, \afe\,$\pm 0.3$), as would be expected given the mean \afe\ of the inner and outer halo samples (\meanafeinner\ and \meanafeouter, respectively).  However, the median \fehsynth\ values for the outer halo fields may be systematically low compared even to the \fehp\ gradient derived using the $\alpha$-enhanced isochrones.

Given the intrinsic field-to-field variation observed in the \fehp\ sample, measurements of \fehsynth\ in additional outer halo fields are needed to confirm that the outer halo \fehsynth\ measurements are indeed systematically lower than the mean \fehp.  Moreover, as noted above, there is significant uncertainty in the magnitude of the systematic bias, due both to failed \fehsynth\ measurements  
and removal of stars with TiO absorption in fields at \rproj\,$\leq$\rprojandone~kpc. 
Recovery of \fehsynth\ for a larger fraction of metal-rich stars in the inner halo fields will also be required before we can confirm significant intrinsic field-to-field variation in the inner halo as well as determine whether the slope of the gradient is consistent between measurements based on \fehsynth\  
and \fehp\@. 
Nevertheless, the current \fehsynth\ data are consistent with a metallicity gradient being present in M31's stellar halo to large radius.

\section{Conclusions}\label{sec:conclusion}

We have presented \feh\ and \afe\ measurements, derived from spectral synthesis, for stars in M31's outer halo spanning a range of projected distance from M31's center of 43 to 165 kpc.  Previous to this work, only four M31 outer halo stars, at projected distances from M31's center of 70 to 140~kpc, had \afe\ or \fehsynth\ abundances \citep{vargas2014ApJL}.  With our \nouterhalofehours\ \fehsynth\ and \nouterhaloafeours\ \afe\ measurements (two of which overlap the \citealt{vargas2014ApJL} sample), we have increased the sample of M31 outer halo stars with \fehsynth\ measurements by a factor of five, and doubled the sample of M31 outer halo \afe\ measurements.  
The final halo sample of \nouterhalofeh\ stars with \fehsynth\ and \nouterhaloafe\ stars with \afe\ measurements were drawn from spectroscopic fields targeting seven of M31's dSph satellites.  

We compared the outer halo star abundances to recent \feh\ and \afe\ measurements of stars in five fields in M31's inner halo \citep[\minrprojinnerhalo\,$\lesssim$\,\rproj\,$\lesssim$\,\maxrprojinnerhalo~kpc;][]{gilbert2019,escala2020}, M31's dwarf satellites \citep{kirby2020,wojno2020}, and the MW\@. 
In contrast to M31's inner halo \citep{escala2020}, the abundances of the M31 outer halo stars are fully consistent with the range of \afe\ vs.\ \fehsynth\ abundances of stars in M31's dSph satellites.  The \afe\ abundances of stars in M31's outer halo
are also more similar to the abundances of MW halo stars \citep[as measured by][]{kirby2010,ishigaki2012,hayes2018} than are the abundances of stars in M31's dynamically hot inner halo.  The stars in the outer halo also have \afe\ patterns that are consistent with the largest of M31's dSph satellites (And~I and And~VII).
A more rigorous comparison of the abundances of stars in M31's outer halo with M31's inner halo, dwarf satellites, and the MW's halo will require a larger sample of outer halo abundances.

We also compared the \fehsynth\ measurements of M31 halo stars to previous CMD-based \fehp\ estimates for a large sample of M31 halo stars drawn from 38 spectroscopic fields \citep[9\,$\lesssim$\,\rproj\,$\lesssim$\,175~kpc;][]{gilbert2014}.  
The \nouterhalofeh\ stars with \fehsynth\ measurements are broadly consistent with both the range and mean of \fehp\ measurements for stars at similar radii in M31's outer halo.  When the mean \fehsynth\ measurements of stars in M31's inner halo are considered, the \fehsynth\ measurements appear to be largely consistent with the metallicity gradient measured from \fehp, albeit with indications of a lower normalization.  However, a robust measurement of the metallicity gradient in M31's stellar halo based on \fehsynth, as well as a determination of the significance level of field-to-field variation in \meanfehsynth\ in M31's stellar halo, will require both larger sample sizes and the recovery of \fehsynth\ for stars with strong TiO absorption.     

The low yield of M31 halo stars at projected distances of $\gtrsim 40$~kpc per observation with current multi-object spectrographs precludes significant near-term progress in measuring \afe\ for a large sample of M31 outer halo stars.  The current sample is based on observations totaling approximately 
45 
hours of exposure time with Keck/DEIMOS.
A similar amount of time spent on dedicated halo masks at these projected distances would produce approximately a factor of two improvement in the number of M31 outer halo stars with \fehsynth\ and \afe\ measurements, since we have found that the yield of M31 halo stars per slitmask is typically comparable in slitmasks targeting dSph and halo fields at large distances \citep[Table~1 of][]{gilbert2012}.  A promising avenue for near-term progress is utilizing the archive of existing Keck/DEIMOS spectroscopic observations to measure mean \fehsynth\ and \afe\ from coadded spectra \citep{wojno2020} from a large number of spectroscopic fields.  Longer-term, wide-field ($\gtrsim 1$ square degree), highly multiplexed (several thousand targets), multi-object spectroscopic facilities on $\gtrsim 8$~m class telescopes have the potential to revolutionize the study of chemical abundances in M31's outer halo \citep[e.g.,][]{takada2014,mse2016,mse2019,gilbert2019astro2020}. 

\acknowledgments
The authors recognize and acknowledge the very significant cultural role and reverence that the summit of Maunakea has always had within the indigenous Hawaiian community. We are most fortunate to have the opportunity to conduct observations from this mountain.  The authors also thank the anonymous referee, whose careful reading of the manuscript improved the clarity of the paper.

Support for this work was provided by NSF grants AST-1614569 (K.M.G., J.W.), AST-1614081 (E.N.K., I.E.), AST-1847909 (E.N.K.), and AST-1909497 (S.R.M.).
P.G. acknowledges support from NSF grants AST-1412648, AST-1010039, AST-0607852, and AST-0307966. S.R.M and R.L.B. acknowledge support from NSF grants AST-1413269, AST-1009882, AST-0607726, and AST-0307842.
E.N.K. gratefully acknowledges support from a Cottrell Scholar award administered by the Research Corporation for Science Advancement as well as funding from generous donors to the California Institute of Technology.  I.E. acknowledges support from a National Science Foundation (NSF) Graduate Research Fellowship under grant No.\ DGE-1745301. 
Support for this work was provided by NASA through Hubble Fellowship grant \#51386.01 awarded to R.L.B. by the Space Telescope Science Institute, which is operated by the Association of  Universities for Research in Astronomy, Inc., for NASA, under contract NAS 5-26555.
The analysis pipeline used to reduce the DEIMOS data was developed at UC Berkeley with support from NSF grant AST-0071048. 

The authors thank A.~McConnachie for use of the PAndAS image.  
This research made use of Astropy, a community-developed core Python package for Astronomy  \citep{astropy2013,astropy2018}\footnote{http://www.astropy.org}. 
  
\facility{Keck:II (DEIMOS).} 
\software{Astropy \citep{astropy2013},
Matplotlib \citep{matplotlib}, numpy \citep{numpy}.}

%%%%%%%%%%%% REFERENCES %%%%%%%%%%%%%%%
\bibliography{m31}

\startlongtable
\begin{splitdeluxetable*}{lrlllrrrRBrrrrRrRr}
\tablecolumns{17}
\tablewidth{0pc}
\tablecaption{Parameters of M31 Halo Stars with Abundance Measurements.\label{table_abund}\tablenotemark{a}}
\tablehead{
\multicolumn{1}{c}{Field} & 
\multicolumn{1}{c}{Object} & 
\multicolumn{2}{c}{Sky Coordinates} & 
\multicolumn{1}{c}{Slitmask} &  
\multicolumn{1}{c}{$R_{\rm proj}$\tablenotemark{b}} &  
\multicolumn{1}{c}{$I_0$} &  
\multicolumn{1}{c}{$(V-I)_0$} &  
\multicolumn{1}{c}{Velocity} &  
\multicolumn{1}{c}{SN} &  
\multicolumn{1}{c}{$T_{\rm eff}$} & 
\multicolumn{1}{c}{$\sigma(T_{\rm eff}$)} &
\multicolumn{1}{c}{log $g$} &
\multicolumn{1}{c}{[Fe/H]} &
\multicolumn{1}{c}{$\sigma$([Fe/H])} &
\multicolumn{1}{c}{[$\alpha$/Fe]} &
\multicolumn{1}{c}{$\sigma$([$\alpha$/Fe])} \\
\multicolumn{1}{c}{Name} &
\multicolumn{1}{c}{Name} &
\multicolumn{1}{c}{RA} & \multicolumn{1}{c}{Dec} & 
\multicolumn{1}{c}{Name} &
\multicolumn{1}{c}{(kpc)} &
\multicolumn{1}{c}{} &
\multicolumn{1}{c}{} &
\multicolumn{1}{c}{(km s$^{-1}$)} &
\multicolumn{1}{c}{(\AA\,$^{-1}$)} &  
\multicolumn{1}{c}{(K)} & 
\multicolumn{1}{c}{(K)} &
\multicolumn{1}{c}{(dex)} &
\multicolumn{1}{c}{(dex)} &
\multicolumn{1}{c}{(dex)} &
\multicolumn{1}{c}{(dex)} &
\multicolumn{1}{c}{(dex)}
} 
\startdata
And I &  7000728 &   00:45:17.07 &   +37:54:36.3 &  and1a &   46.5 &   21.2 &    1.1 &  -303.7 &  20.2 & 4597 & 50  & 1.03 &  -1.44 &  0.13 &  0.03 &  0.36 \\
And I &  7000611 &   00:45:18.22 &   +37:55:59.4 &  and1a &   46.2 &   21.4 &    1.3 &  -156.7 &  13.5 & 4291 & 33  & 0.98 &  -1.74 &  0.16 &  0.94 &  0.4 \\
And I &  6000883 &   00:45:19.80 &   +37:58:28.7 &  and1a &   45.6 &   21.8 &    1.9 &  -373.5 &  10.0 & 3794 & 21  & 0.94 &  -0.37 &  0.13 &  ...  &  ...  \\
And I &  7000018 &   00:45:24.58 &   +37:55:30.8 &  and1a &   46.3 &   22.0 &    1.4 &  -436.9 &  8.7 & 4167 & 34  &  1.17 & -1.23 &  0.15 &  ...  &  ...  \\ 
And I     &  1007290   &  00:45:46.98 &  +38:11:28.0 &  d1\_1              &   42.9 &   21.8 &    1.2 &  -435.2 &    5.4 &    4467 &     53 &   1.25 &  -1.73 &   0.29 &    ... &    ... \\  
And I     &  1007432   &  00:45:43.67 &  +38:11:55.7 &  d1\_1              &   42.8 &   21.9 &    1.4 &  -419.9 &    3.5 &    4121 &     42 &   1.13 &  -0.86 &   0.21 &    ... &    ... \\  
And I     &  2006273   &  00:46:41.27 &  +38:03:3.6  &  d1\_2              &   45.3 &   20.5 &    1.3 &  -439.5 &   10.6 &    4313 &     10 &   0.65 &  -1.51 &   0.10 &  -0.24 &   0.32 \\  
And III &  3007800 &   00:35:29.64 &   +36:25:02.3 &  and3a &   69.4 &   22.3 &    1.2 &  -411.3 &  9.8 & 4420 & 58  & 1.43 &  -1.61 &  0.18 &  ...  &  ...  \\
And III &  8000203 &   00:35:12.60 &   +36:21:07.5 &  and3a &   70.5 &   22.3 &    1.0 &  -432.0 &  7.6 & 4897 & 74  & 1.58 &  -0.63 &  0.15 &  ... &  ...  \\
And III  &  4002365   &  00:35:30.57 &  +36:19:15.8 &  d3\_2              &   70.7 &   20.6 &    1.3 &  -329.0 &   15.7 &    4330 &      6 &   0.70 &  -1.73 &   0.09 &   0.24 &   0.21 \\
And III  &  4002335   &  00:35:39.72 &  +36:21:9.3  &  d3\_2              &   70.1 &   21.0 &    1.3 &  -360.7 &   12.6 &    4231 &     10 &   0.79 &  -2.31 &   0.15 &   0.13 &   0.34 \\
And III  &  3006203   &  00:35:39.14 &  +36:24:37.3 &  d3\_2              &   69.4 &   21.7 &    1.3 &  -288.6 &    5.1 &    4300 &     26 &   1.11 &  -1.47 &   0.17 &   ...  &   ...  \\
And III  &  4001981   &  00:35:30.76 &  +36:15:31.2 &  d3\_2              &   71.5 &   21.9 &    1.4 &  -304.6 &    5.0 &    4118 &     28 &   1.15 &  -1.54 &   0.25 &   ...  &   ...  \\
And III  &  2001738   &  00:36:0.41  &  +36:37:5.4  &  d3\_3              &   66.3 &   21.7 &    1.3 &  -306.8 &    4.1 &    4235 &     35 &   1.11 &  -0.97 &   0.17 &   ...  &   ...  \\
And XV   &  2070      &  01:15:1.53  &  +38:05:21.4 &  d15\_1             &   96.3 &   21.3 &    1.1 &  -303.8 &   11.2 &    4669 &     50 &   1.07 &  -2.65 &   0.39 &    ... &    ... \\
And XV   &  497       &  01:14:58.26 &  +38:00:59.1 &  d15\_2             &   96.7 &   21.1 &    1.1 &  -408.5 &   14.1 &    4687 &     41 &   1.02 &  -1.21 &   0.09 &   0.10 &   0.34 \\
And V &  3007277 &   01:09:56.51 &   +47:37:49.8 &  and5b &  110.7 &   21.5 &    1.5 &  -213.8 &  15.6 & 4041 & 29  & 0.90 &  -1.34 &  0.12 &  ...  &  ...  \\
And V &  3007616 &   01:10:02.05 &   +47:36:25.5 &  and5b &  110.6 &   22.4 &    1.1 &  -268.4 &  8.0 & 4482 & 88  & 1.45 &  -2.09 &  0.33 &  ...  &  ...  \\
And V &  3008720 &   01:09:54.41 &   +47:35:18.3 &  and5b &  110.2 &   20.7 &    1.4 &  -221.7 &  30.6 & 4143 & 24  & 0.61 &  -2.95 &  0.15 &  0.98 &  0.18 \\
And V    &  2006635   &  01:10:6.50  &  +47:42:41.2 &  d5\_1              &  111.9 &   21.3 &    1.2 &  -305.9 &    8.2 &    4521 &     29 &   1.01 &  -2.18 &   0.28 &    ... &    ... \\   
And XIV  &  5007496   &  00:51:28.10 &  +29:42:10.6 &  d14\_3             &  164.5 &   21.2 &    1.1 &  -248.4 &    6.9 &    4726 &     27 &   1.10 &  -1.53 &   0.29 &  ... 0 &   ...  \\
\enddata
\tablenotetext{a}{Measurements of \feh\ and \afe\ included here pass all quality and sample selection criteria discussed in Section~\ref{sec:data}.  
}
\tablenotetext{b}{The projected distance of the star from the center of M31 in the tangent plane, assuming a distance modulus of $24^{\rm m}.47$ \citep{stanek1998, mcconnachie2005}.
}
\end{splitdeluxetable*}

\end{document}